\documentclass[english,aps,a4paper,prd,groupedaddress,preprintnumbers,nofootinbib,preprint,floatfix,showpacs]{revtex4}

\usepackage{hyperref}
\usepackage{amsmath,bbm,latexsym,amssymb}
\usepackage{graphicx}
\usepackage{rotating}
\usepackage{color} 
\usepackage{multirow} 
\usepackage[T1]{fontenc} 
\usepackage[utf8]{inputenc}
\usepackage{lmodern}
\makeatletter

\usepackage{booktabs}
\usepackage{mathrsfs}   
\usepackage{slashed}     
\usepackage{url}
\usepackage{multirow}
\usepackage{units}
\usepackage{wasysym }
\usepackage{caption}
\usepackage{comment}

\usepackage{float}
\newcommand{\bea}{\begin{eqnarray}}
\newcommand{\eea}{\end{eqnarray}}

\newcommand{\be}{\begin{equation}}
\newcommand{\ee}{\end{equation}}
\def\beq{\begin{equation}}
\def\eeq{\end{equation}}
\newcommand{\ba}{\begin{eqnarray}}
\newcommand{\ea}{\end{eqnarray}}
\def\half{\ifmath{{\textstyle{\frac{1}{2}}}}}

 \def\ifmath#1{\relax\ifmmode #1\else $#1$\fi}
\def\eq#1{Eq.~(\ref{#1})}
\def\eqs#1#2{Eqs.~(\ref{#1}) and (\ref{#2})}

\def\sgi  {s_{\gamma_1}}
\def\cgi  {c_{\gamma_1}}
\def\sgii  {s_{\gamma_2}}
\def\cgii  {c_{\gamma_2}}
\def\sai  {s_{\alpha_1}}
\def\cai  {c_{\alpha_1}}
\def\saii  {s_{\alpha_2}}
\def\caii  {c_{\alpha_2}}
\def\saiii  {s_{\alpha_3}}
\def\caiii  {c_{\alpha_3}}

\makeatother

\usepackage{babel}

\newcommand{\Z}[1]{\ensuremath{\mathbbm{Z}_{#1}}} 

\def\vb#1{\vbox to #1 pt{}}

\makeatletter
\renewcommand*\env@matrix[1][\arraystretch]{%
  \edef\arraystretch{#1}%
  \hskip -\arraycolsep
  \let\@ifnextchar\new@ifnextchar
  \array{*\c@MaxMatrixCols c}}
\makeatother

\newcommand{\AddrCFTP}{%
 Departamento de F\'\i sica and CFTP, Instituto Superior T\'ecnico\\
 Universidade de Lisboa, 
          Av. Rovisco Pais 1, 1049-001 Lisboa, Portugal }

\def\gsim{\raise0.3ex\hbox{$\;>$\kern-0.75em\raise-1.1ex\hbox{$\sim\;$}}}
\def\lsim{\raise0.3ex\hbox{$\;<$\kern-0.75em\raise-1.1ex\hbox{$\sim\;$}}}

\allowdisplaybreaks

\begin{document}

\preprint{CFTP/21-010}

\title{Current bounds on the Type-Z $Z_3$ three Higgs doublet model}
\author{Rafael Boto}%
\email{rafael.boto@tecnico.ulisboa.pt}
\affiliation{\AddrCFTP}
\author{Jorge C.~Rom\~ao}\email{jorge.romao@tecnico.ulisboa.pt}
\affiliation{\AddrCFTP} 
 \author{João  P.~Silva} \email{jpsilva@cftp.ist.utl.pt}
 \affiliation{\AddrCFTP}

\today

\begin{abstract}
Type-Z models, where charged leptons, up type quarks and down type quarks
each couple to a different scalar, are only possible when there are three
or more Higgs doublets.
We consider the Type-Z three Higgs doublet model imposed by a
softly broken $\Z3$ symmetry.
We take into account all theoretical and experimental constraints,
including perturbative unitarity
and bounded from below conditions that we develop here.
Since there can be cancellations between the two charged
Higgs in $B \rightarrow X_s \gamma$ (and in $h \rightarrow \gamma \gamma$),
the lower bounds obtained on the charged Higgs masses are alleviated.
We also discuss in detail the important physical differences between
exact alignment and approximate alignment,
and present some useful benchmark points.
\end{abstract}

\maketitle

After the observation in 2012 by ATLAS and CMS
\cite{Aad:2012tfa,Chatrchyan:2012ufa} of a new scalar particle closely
resembling the Standard Model (SM) Higgs boson, the search for physics
beyond the Standard Model (BSM) is now the main goal of the LHC
experiments. Popular extensions where only Higgs doublets are added to
the SM have been extensively studied and allow for both the agreement
with the experimental results and the possibility of  new features;
for reviews see \cite{Gunion:1989we,Branco:2011iw,Ivanov:2017dad}.   

The simplest extension, the two-Higgs-doublet model (2HDM),
can provide new sources of CP-violation necessary to fulfill
the Sakharov criteria for baryogenesis \cite{Sakharov:1967dj}.
However, the most general Higgs-fermion Yukawa
couplings generically yield Higgs-mediated flavor-changing neutral
``currents'' (FCNCs) at tree level,
in conflict with experimental observations.
A common method to have FCNCs sufficiently suppressed is to impose
symmetries on the Lagrangian:  tree-level FCNC effects can be
completely removed by establishing how the fermion and scalar fields
have to transform under the chosen symmetry.  
In the two Higgs doublet model (2HDM) this can be achieved
by imposing a $\Z2$ symmetry \cite{Glashow:1976nt,Paschos:1976ay}.
Reference~\cite{Ferreira:2010xe} showed  that in general N Higgs doublet models (NHDM) the Yukawa coupling matrices
to fermions of a given electric charge remain proportional (thus removing
FCNCs) under the renormalization group running if and only if there is a
basis for the Higgs doublets in which all the fermions of a given
electric charge couple to only one Higgs doublet.
The models are then classified based on these choices.
The four (five) distinct types of Yukawa couplings in models with two
(more than two) doublets that fit this requirement were introduced in
\cite{Ferreira:2010xe} and denoted in \cite{Yagyu:2016whx} 
by Types I, II, X (also known as lepton-specific), Y (flipped), and Z,
according to 
\begin{eqnarray}
\textrm{Type-I:}
&&
\phi_u=\phi_d=\phi_e\, ,
\nonumber\\
\textrm{Type-II:}
&&
\phi_u \neq \phi_d=\phi_e\, ,
\nonumber\\
\textrm{Type-X:}
&&
\phi_u=\phi_d \neq \phi_e\, ,
\nonumber\\
\textrm{Type-Y:}
&&
\phi_u=\phi_e \neq \phi_d\, ,
\nonumber\\
\textrm{Type-Z:}
&&
\phi_u\neq \phi_d;\ \phi_d\neq \phi_e,\ \phi_e\neq \phi_u\, ,
\end{eqnarray}
where $\phi_{u,d,e}$ are the single scalar fields that couple
exclusively to the up type quarks, down type quarks, and charged
leptons, respectively.  In this work, we set our attention on the
Type-Z that can only appear for NHDM with $N>2$. It is interesting to
see what differences there are in this new type of model, since it
decouples completely the up quark, down quark and charged lepton
sectors from one-another.

There are have been implementations of Type-Z in three-Higgs-doublet
models (3HDM) using a $\Z2\times\Z2$ symmetry
\cite{Akeroyd:2020nfj,Logan:2020mdz} or $\Z3$
\cite{Das:2019yad,Boto:2021}.  For this work, we choose to use a $\Z3$
symmetric potential.  This symmetry is realizable through the
following representation,
\begin{equation}
    S_{\Z3}=\text{diag}(1,e^{i\frac{2\pi}{3}}e^{-i\frac{2\pi}{3}}) .
\end{equation}
Recently, there has been an analysis of $\Z3$ 3HDM which takes the
exact alignment limit and looks at specific values of the physical
parameters \cite{Chakraborti:2021bpy}.  It does not seem to consider
the theoretical constraints coming from perturbative unitarity,
discussed explicitly for the $\Z3$ 3HDM model in
Ref.~\cite{Bento:2017eti} and BFB conditions, which we develop here.
Compatibility with the bounds coming from Higgs searches is also
checked with the newest version of the \texttt{HiggsBounds-5.9.1}
(HB5) code \cite{Bechtle:2020pkv}.  In particular, we show that recent
LHC measurements exclude all points in Fig.~2 of
Ref.~\cite{Chakraborti:2021bpy}, for the same parameter choices.  We
then show that by scanning for a larger range of parameters (away from
exact alignment, but still consistent with all experimental data) we
can obtain viable points corresponding to smaller masses for the
additional particles.

In Sec.~\ref{sect:model} we describe succinctly the scalar and Yukawa
sectors of the $\Z3$ 3HDM model,
discussed also in \cite{Das:2019yad,Boto:2021,Chakraborti:2021bpy}.
The theoretical and experimental constraints are described
in Sec.~\ref{sect:constraints}.
In Sec.~\ref{sec:hgaga} we describe the impact of current LHC measurements
on the 125GeV scalar decays, both excluding and including the impact
of HB5 bounds.
In particular, we discuss the fact that the couplings of the 
two charged scalars may have different signs,
thus allowing for canceling contributions to $h \rightarrow \gamma \gamma$.
A similar effect is possible in $B\to X_s \gamma$,
thus alleviating the lower bounds on charged scalar masses.
This is discussed in Sec.~\ref{sect:xsgamma} and Sec.~\ref{sec:Fig2},
where we explore the regions of parameters allowed by the different
constraints imposed,
starting from the experimental limits on the BR($B\to X_s \gamma$)
and progressively varying the ranges on our parameter scans.
Our work highlights the importance of going beyond strict alignment,
when procuring the full range of possibilities existent within
the $\Z3$ 3HDM.
We present illustrative benchmark points in Sec.~\ref{sec:BP}
and
discuss our conclusions in Sec.~\ref{sec:conclusions},
leaving the appendix for the full expression of some couplings
required in our calculations.

\section{\label{sect:model}The $\Z3$ 3HDM Model}
 
\subsection{\label{subsec:scalar}Scalar sector} 
 
Taking the potential defined by \cite{Das:2019yad}, the terms
invariant under the chosen transformation, $\phi_i\to
\phi_i'=(S_{\Z3})_{ij}\phi_j$, are given by 
\begin{equation}\label{Z3potential}
    V_{\Z3}=V_{\text{quadratic}}+V_{\text{quartic}},
\end{equation}
with the quartic part
\ba
V_{\text{quartic}}&=&
\lambda_1(\phi_1^\dagger\phi_1)^2
+\lambda_2(\phi_2^\dagger\phi_2)^2
+\lambda_3(\phi_3^\dagger\phi_3)^2+\lambda_4(\phi_1^\dagger\phi_1)(\phi_2^\dagger\phi_2)+\lambda_5(\phi_1^\dagger\phi_1)(\phi_3^\dagger\phi_3)\nonumber\\[8pt]
&&\quad 
+\lambda_6(\phi_2^\dagger\phi_2)(\phi_3^\dagger\phi_3)+\lambda_7(\phi_1^\dagger\phi_2)(\phi_2^\dagger\phi_1)+\lambda_8(\phi_1^\dagger\phi_3)(\phi_3^\dagger\phi_1)
+\lambda_9(\phi_2^\dagger\phi_3)(\phi_3^\dagger\phi_2)\nonumber\\[8pt]
&&\quad 
+\left[\lambda_{10}(\phi_1^\dagger\phi_2)(\phi_1^\dagger\phi_3)
+\lambda_{11}(\phi_1^\dagger\phi_2)(\phi_3^\dagger\phi_2)
+\lambda_{12}(\phi_1^\dagger\phi_3)(\phi_2^\dagger\phi_3)+\text{h.c.}\right],\label{Z3quartic}
\ea
and the quadratic part 
\begin{equation}
    V_{\text{quadratic}}=m_{11}^2\phi_1^\dagger\phi_1+m_{22}^2\phi_2^\dagger\phi_2+m_{33}^2\phi_3^\dagger\phi_3 +\left[m_{12}^2(\phi_1^\dagger\phi_2)
+m_{13}^2(\phi_1^\dagger\phi_3)
+m_{23}^2(\phi_2^\dagger\phi_3)+\text{h.c.}\right] ,
\end{equation}
also including terms, $m_{12}^2$, $m_{13}^2$ and $m_{23}^2$,
that break the symmetry softly.

After spontaneous symmetry breaking (SSB), the three doublets can be parameterized in terms of its component fields as:\footnote{Notice that we use
$x_i$ in place of Ref.~\cite{Das:2019yad}'s $h_i$,
because for us $h_i$ are the physical neutral scalar mass eigenstates.}
 \begin{equation}
     \phi_i=\begin{pmatrix} w_k^\dagger \\ (v_i+x_i\,+\,i\,z_i)/\sqrt{2}\end{pmatrix} \,\,,\qquad (i=1,2,3)\label{fielddefinitions}
 \end{equation}
where $v_i/\sqrt{2}$ corresponds to the vacuum expectation value
(vev) for the neutral component of $\phi_i$.
It is assumed that the
scalar sector of the
model explicitly and spontaneously conserves
CP.\footnote{Strictly speaking,
it is not advisable to assume a real scalar sector while allowing the
Yukawa couplings to carry the phase necessary for the
CKM matrix.
This is also a problem with the so-called real 2HDM
\cite{Fontes:2021znm}.
One can take the view that the complex terms and their counterterms
in the scalar sector exist, with the former set to zero.
}

That is, all the parameters in the scalar potential are real and
the vevs $v_1$, $v_2$ , $v_3$, are also real.
With this assumption, the scalar potential of \eq{Z3potential}
contains eighteen parameters. The vevs can be parameterized as follows:
\begin{equation}\label{3hdmvevs}
     v_1=v \cos \beta_1 \cos \beta_2\,,\qquad v_2=v \sin \beta_1 \cos \beta_2\, ,\qquad v_3=v \sin \beta_2,
\end{equation}
leading to the Higgs basis  \cite{Georgi:1978ri,Donoghue:1978cj,Botella:1994cs}
to be obtained by the following rotation,
\begin{equation}\label{higgsbasisZ3}
     \begin{pmatrix} H_0 \\ R_1 \\ R_2 \end{pmatrix}
     =
     \mathcal{O}_\beta
     \begin{pmatrix} x_1 \\ x_2 \\ x_3 \end{pmatrix}
     =
     \begin{pmatrix} \cos\beta_2 \cos\beta_1 & \cos\beta_2 \sin\beta_1 & \sin\beta_2 \\ -\sin\beta_1 & \cos\beta_1 & 0 \\ -\cos\beta_1 \sin\beta_2 & -\sin\beta_1 \sin\beta_2 & \cos\beta_2\end{pmatrix}
     \begin{pmatrix} x_1 \\ x_2 \\ x_3 \end{pmatrix} .
\end{equation}
The scalar kinetic Lagrangian is written as
\begin{equation}\label{kinetic3hdm}
    \mathscr{L}_{\text{kin}}=\sum_{k=1}^{n=3}|D_\mu\phi_k|^2 ,
\end{equation}
and contains the terms relevant to the propagators and trilinear couplings of the scalars and gauge bosons.

 We can now define orthogonal matrices which diagonalize the
 squared-mass matrices present in the CP-even scalar, CP-odd scalar
 and charged scalar sectors. These are the transformations that take
 us to the physical basis, with states possessing well-defined
 masses. Following Ref.~\cite{Das:2019yad,Boto:2021}, the twelve
 quartic couplings can be exchanged for seven physical masses (three 
CP-even scalars, two CP-odd scalars and two pairs of charged scalars)
and five mixing angles. The mass terms in the neutral scalar sector
can be extracted through the following rotation,  
\begin{equation}\label{CPevenDiag}
     \begin{pmatrix} h_1 \\ h_2 \\ h_3 \end{pmatrix}=\mathcal{O}_\alpha \begin{pmatrix} x_1 \\ x_2 \\ x_3 \end{pmatrix},
\end{equation}
where we take $h_1 \equiv h_{125}$ to the be the 125GeV Higgs particle
found at LHC.
The form chosen for $\mathcal{O}_\alpha$ is
\begin{equation}\label{matrixR}
\textbf{R}\equiv\mathcal{O}_\alpha=\mathcal{R}_3.\mathcal{R}_2.\mathcal{R}_1 ,
\end{equation}
where
\begin{equation}
\mathcal{R}_1=\begin{pmatrix}\cai & \sai & 0\\ -\sai & \cai & 0 \\ 0 & 0 & 1  \end{pmatrix}\,,\quad     \mathcal{R}_2=\begin{pmatrix}\caii & 0 & \saii \\ 0 & 1 & 0 \\ -\saii & 0 & \caii  \end{pmatrix}\,,\quad     \mathcal{R}_3=\begin{pmatrix}1 & 0 & 0\\ 0 & \caiii & \saiii \\ 0 & -\saiii & \caiii  \end{pmatrix}\,.\quad
\end{equation}

  For the CP-odd scalar sector, the physical basis is chosen as $\begin{pmatrix}G^0 & A_1 & A_2\end{pmatrix}^T$ and the transformation to be
   \begin{equation}\label{CPoddDiag}
     \begin{pmatrix} G^0 \\ A_1 \\ A_2 \end{pmatrix}=\mathcal{O}_{\gamma_1} \mathcal{O}_\beta \begin{pmatrix} z_1 \\ z_2 \\ z_3 \end{pmatrix} ,
 \end{equation}
where $\mathcal{O}_{\gamma_1}$ is defined in order to diagonalize the 2x2 submatrix that remains in the Higgs basis, with the form
  \begin{equation}
     \mathcal{O}_{\gamma_1}= \begin{pmatrix}1  & 0 & 0\\ 0& \cgi & -\sgi \\ 0 & \sgi & \cgi \end{pmatrix} . \label{ogamma1}
  \end{equation}
For later use, we define the matrix $\textbf{P}$ as the combination
\begin{equation}\label{matrixP}
    \textbf{P}\equiv\mathcal{O}_{\gamma_1} \mathcal{O}_\beta .
\end{equation}

For the charged scalar sector, the physical basis is 
$\begin{pmatrix}G^+ & H_1^+ & H_2^+\end{pmatrix}^T$
and the transformation is
\begin{equation}\label{ChargedDiag}
\begin{pmatrix} G^+ \\ H_1^+ \\ H_2^+ \end{pmatrix}
=\mathcal{O}_{\gamma_2} \mathcal{O}_\beta
\begin{pmatrix} w_1^\dagger \\ w_2^\dagger \\ w_3^\dagger \end{pmatrix},
\end{equation}
where
\begin{equation}
\mathcal{O}_{\gamma_2}= \begin{pmatrix}1  & 0 & 0\\ 0& \cgii & -\sgii \\ 0 & \sgii & \cgii \end{pmatrix} .
\label{ogamma2} 
\end{equation}
We write the masses of $H_1^+$ and $H_2^+$
as $m_{H_1^\pm}$ and $m_{H_2^\pm}$, respectively.
The matrix $\textbf{Q}$ is then defined as the combination 
\begin{equation}\label{matrixQ}
\textbf{Q}\equiv\mathcal{O}_{\gamma_2} \mathcal{O}_\beta .
\end{equation}

Considering that the states in the physical basis have well-defined masses, we can obtain relations between the set 
\ba
&&\left\{v_1,v_2,v_3,m_{h1},m_{h2},m_{h3},m_{A1},m_{A2},m_{H_1^\pm},m_{H_2^\pm},\alpha_1,\alpha_2,\alpha_3,\gamma_1,\gamma_2\right\} ,\label{setphysical}\\[8pt]
&&\quad 
v_1=v \cos\beta_1 \cos\beta_2\,,\quad\,v_2=v \sin\beta_1 \cos\beta_2\,,\quad\,v_3=v \sin\beta_2 ,
\ea
and the parameters of the potential in \eq{Z3potential},
as shown in Ref.~\cite{Das:2019yad,Boto:2021}.
We performed an extensive scan of the parameter space in \eq{setphysical}.
Our fixed inputs are $v = 246\,\text{GeV}$ and $m_{h1} = 125\,\text{GeV}$.
We then took random values in the ranges:
\ba
&&\alpha_1,\, \alpha_2,\, \alpha_3,\, \gamma_1,\, \gamma_2\, \in \left[-\frac{\pi}{2},\frac{\pi}{2}\right];\qquad \tan{\beta_1},\,\tan{\beta_2}\,\in \left[0,10\right];
\nonumber\\[8pt]
&&m_{h2},\,m_{h3}\,\in \left[125,1000\right]\,\text{GeV};\qquad m_{A_1},\,m_{A_2}\,m_{H_1^\pm},\,m_{H_2^\pm}\,\in \left[100,1000\right]\,\text{GeV}.
\label{scanparameters}
\ea
These parameter ranges will be used in all scans and figures
presented below, except where noted otherwise.
The lower limits chosen for the masses satisfy the
constraints listed in
Ref.~\cite{Aranda:2019vda}.\footnote{Ref.~\cite{Aranda:2019vda} has
the same $\Z3$ 3HDM scalar sector,
but it does not couple to fermions as a Type-Z model
because the aim there is to have two Inert scalar doublets and only one
active one.}

 \subsection{Higgs-Fermion Yukawa interactions}

One can now impose the Type-Z model on the Yukawa Lagrangian, by
establishing how the fields behave under the $\Z3$ transformation. For
this, there are multiple possibilities that differ on which of the
scalars gives mass to each type of fermion. We follow the choice made
by Das and Saha \cite{Das:2019yad}. The scalar doublets $\phi_1$ and
$\phi_2$ transform nontrivially as: 
\begin{equation}\label{3hdmPhiCharge}
    \phi_1\to\omega\phi_1\,,\qquad\phi_2\to\omega^2\phi_2,
\end{equation}
where $\omega=e^{2\pi\,i/3}$. For the fermionic fields, we consider that under $\Z3$
\begin{equation}\label{3hdmFermCharge}
    d_R\to\omega d_R\,,\qquad l_R\to\omega^2\,l_R ,
\end{equation}
while the rest of the fields remain unaffected. It follows that the
Yukawa coupling matrices are now restricted: $\phi_1$ only has
interaction terms with the charged leptons, giving them mass; $\phi_3$
and $\phi_2$ are responsible for masses of the up and down type
quarks, respectively.   

When taking into account the restrictions imposed by the symmetry, the
Yukawa couplings to fermions can be written in a compact form. For the
couplings of neutral Higgs to fermions, 
\begin{equation}\label{couplingNeutralFerm}
    \mathscr{L}_{\rm Y}\ni -\frac{m_f}{v}\bar{f}(a^f_j+i\, b^f_j\gamma_5)fh_j ,
\end{equation}
where we group the physical Higgs fields in a vector, as $h_j\equiv(h_1,h_2,h_3,A_1,A_2)_j$. The coefficients are given in \eq{coeffNeutralFerm},
\ba
a_j^f &\to&
\frac{\textbf{R}_{j,1}}{\hat{v_1}},
\qquad\qquad j=1,2,3\qquad \text{for all leptons} ,\nonumber\\[8pt]
b_j^f &\to&
\frac{\textbf{P}_{j-2,1}}{\hat{v_1}},
\qquad\quad j=4,5\quad\qquad \text{for all leptons} ,\nonumber\\[8pt]
a_j^f &\to&
\frac{\textbf{R}_{j,3}}{\hat{v_3}},
\qquad\qquad j=1,2,3\qquad \text{for all up quarks} ,\nonumber\\[8pt]
b_j^f &\to&
-\frac{\textbf{P}_{j-2,3}}{\hat{v_3}},
\quad\quad j=4,5\quad\qquad \text{for all up quarks} ,\nonumber\\[8pt]
a_j^f &\to&
\frac{\textbf{R}_{j,2}}{\hat{v_2}},
\qquad\qquad j=1,2,3\qquad \text{for all down quarks} ,\nonumber\\[8pt]
b_j^f &\to&
\frac{\textbf{P}_{j-2,2}}{\hat{v_2}},
\qquad\quad j=4,5\quad\qquad \text{for all down quarks} ,
\label{coeffNeutralFerm}
\ea
where we introduce $\hat{v_i}=v_i/v$, with the vevs in \eq{3hdmvevs}.
Note how the coupling of each type of fermion depends on entries of the diagonalization matrices in \eqs{matrixR}{matrixP}. 

The couplings of the charged Higgs, $H_1^\dagger$ and $H_2^\dagger$, to fermions can be expressed as
\begin{eqnarray}
\mathscr{L}_{\rm Y} &\ni& \frac{\sqrt{2}}{v}
\bar{\psi}_{d_i}
\left[m_{\psi_{d_i}} V_{ji}^\ast\, \eta_k^L P_L
+ m_{\psi_{u_j}} V_{ji}^\ast\, \eta_k^R P_R\right] \psi_{u_j} H_k^-
\nonumber\\
&&
+ \frac{\sqrt{2}}{v}\bar{\psi}_{u_i}
\left[m_{\psi_{d_j}} V_{ij}\, \eta_k^L P_R 
+ m_{\psi_{u_i}} V_{ij}\, \eta_k^R P_L \right] \psi_{d_j} H_k^+ ,
\label{couplingChargedFerm}
\end{eqnarray}
where $(\psi_{u_i},\psi_{d_i})$ is $(u_i,d_i)$ for quarks or $(\nu_i,l_i)$ for leptons.
For quarks, $V$ is the CKM matrix, while for leptons,
$V_{ij}=\delta_{ij}$ since we are considering massless neutrinos.
The couplings are
\begin{equation}
    \eta_k^{l\,L}=-\frac{\textbf{Q}_{k+1,1}}{\hat{v_1}}\,,\quad\eta_k^{l\,R}=0\,,\quad\eta_k^{q\,L}=-\frac{\textbf{Q}_{k+1,2}}{\hat{v_2}}\,,\quad\eta_k^{q\,R}=\frac{\textbf{Q}_{k+1,3}}{\hat{v_3}}\,,\quad \text{k=1,2} ,
\end{equation}
for leptons and quarks, respectively.

\section{\label{sect:constraints}Constraints on the parameter space}

In this section we study the constraints that must be applied to the
model parameters in order to ensure consistency. 

\subsection{Theoretical Constraints 1}

We impose perturbativity unitarity, sufficient bounded from below (BFB)
conditions, and the oblique parameters $S$, $T$, and $U$.

\subsubsection{BFB conditions on the 3HDM}
As basic requirements for any physical theory, the Higgs potential
must satisfy conditions that ensure it  possesses a stable minimum,
around which one can perform perturbative calculations.  That is, it
must be bounded from below, meaning that there is no direction in
field space along which the value of the potential tends to minus
infinity.  
This need of a non-trivial minimum is then translated to conditions
on the parameters of the potential.

Focusing on the study of the 3HDM constrained by a  $\Z3$ symmetry, the quartic terms in \eq{Z3quartic} can be written as
\begin{equation}\label{VquarticSep}
    V_{\text{quartic}}=V_0+V_1 ,
\end{equation}
where $V_0$ has the terms in $\lambda_{1\to 9}$ and $V_1$ the terms $\lambda_{10\to12}$.
If the potential were just $V_0$ in \eq{VquarticSep},
then the BFB necessary and sufficient conditions would be
simply those given by Klimenko in Ref.~\cite{Klimenko:1984qx}.
The problem, not yet solved for the 3HDM
with a $\Z3$ symmetry is the $V_1$ part.
We will introduce sufficient conditions for BFB by bounding the
potential by a lower potential.
To do that we follow \cite{Klimenko:1984qx,Fontes:2019uld},
checking for neutral minima.
Neutral directions in the Higgs space correspond to situations when all
$\phi_i$ are proportional to each
other\footnote{Other directions, along which the strict
proportionality of all three doublets does not hold,
are called \textit{charge-breaking} (CB) directions.
In recent works \cite{Faro:2019vcd,Ivanov:2020jra},
it has been proven that these directions can lead to
pathological situations for other symmetries in the 3HDM.
It is then required to consider these directions when doing
a complete work of looking for necessary and sufficient BFB conditions.
Our contribution to the analysis of the $\Z3$ symmetry is to
specify sufficient conditions along the neutral direction.}.
Along these directions, we can then define
\begin{equation}
    \phi_1\to\sqrt{x}e^{i\theta_1},\qquad \phi_2\to\sqrt{y}e^{i\theta_2},\qquad \phi_3\to\sqrt{z}e^{i\theta_3} .
\end{equation}
It then follows that for $V_0$,
\ba
V_0&=&\lambda_1 x^2 +\lambda_2 y^2 + \lambda_3 z^2 + \lambda_4 xy +\lambda_5 xz +\lambda_6 yz + \lambda_7 xy + \lambda_8 xz + \lambda_9 yz \nonumber\\[8pt]
&=& \lambda_1 x^2 +\lambda_2 y^2 + \lambda_3 z^2 + (\lambda_4+\lambda_7) xy +(\lambda_5+\lambda_8) xz +(\lambda_6+\lambda_9) yz ,\label{V0final}
\ea
and for $V_1$,
\begin{equation}
    V_1=2\lambda_{10}x\sqrt{y}\sqrt{z}\cos \delta_1 + 2\lambda_{11}y\sqrt{x}\sqrt{z}\cos \delta_2 + 2 \lambda_{12}z\sqrt{x}\sqrt{y}\cos \delta_3 ,
\end{equation}
where $\delta_i$ are some combination of the phases $\theta_i$. Considering that $x,y,z > 0$ by definition, we can start our strategy of bounding the potential by a lower one with
\begin{equation}
    V_1 \geq V_1'= -2|\lambda_{10}|x\sqrt{y}\sqrt{z}-2|\lambda_{11}|y\sqrt{x}\sqrt{z}-2|\lambda_{12}|z\sqrt{x}\sqrt{y} .
\end{equation}
Notice that for non-negative $x,y,z$ one has 
\begin{equation}
    -\sqrt{x}\sqrt{z}>-x-y,\qquad - \sqrt{x} \sqrt{z}> -x-z,\qquad -\sqrt{y} \sqrt{z}>-y-z.
\end{equation}
Therefore,
\begin{equation}
V_1\geq V_1' > V_1''=-2|\lambda_{10}|(xy+xz)-2|\lambda_{11}|(xy+yz)-2|\lambda_{12}|(xz+yz) ,\label{V1final}
\end{equation}
and combining \eq{V1final} with \eq{V0final}, it follows that
\begin{equation}
    V_0+V_1>V_{\text{BFB}},
\end{equation}
where
\begin{equation}
    V_{\text{BFB}}=\lambda_1x^2+\lambda_2y^2+\lambda_3z^2+2\alpha x y + 2\beta x z + 2 \gamma y z,
\end{equation}
with the definitions,
\ba
\alpha&=&\half (\lambda_4+\lambda_7-2|\lambda_{10}|-2|\lambda_{11}|) ,\nonumber\\[8pt]
\beta&=&\half (\lambda_5+\lambda_8-2|\lambda_{10}|-2|\lambda_{12}|) ,\nonumber\\[8pt]
\gamma&=&\half (\lambda_6+\lambda_9-2|\lambda_{11}|-2|\lambda_{12}|) .
\ea

Now, for the potential $V_{\text{BFB}}$ the necessary and sufficient
conditions are obtained from Ref.~\cite{Klimenko:1984qx}:
\begin{itemize}
\item $\lambda_1>0,\lambda_2>0,\lambda_3>0,$
\item $\left\{\beta>-\sqrt{\lambda_1\lambda_3}; \gamma>-\sqrt{\lambda_2\lambda_3}; \alpha>-\sqrt{\lambda_1\lambda_2}; \beta\geq -\gamma \sqrt{\lambda_1/\lambda_2}\right\}$\begin{equation}
        \cup\left\{\sqrt{\lambda_2\lambda_3}>\gamma>-\sqrt{\lambda_2\lambda_3};\quad-\gamma\sqrt{\lambda_1/\lambda_2}\geq\beta > -\sqrt{\lambda_1\lambda_3};\quad \lambda_3\alpha>\beta\gamma - \sqrt{\Delta_\alpha \Delta_\gamma}\right\} ,
    \end{equation}
\end{itemize}
where
\begin{equation}
\Delta_\alpha=\beta^2-\lambda_1\lambda_3,
\quad \Delta_\gamma=\gamma^2-\lambda_2\lambda_3.
\end{equation}
As $V_0+V_1>V_{\text{BFB}}$, these conditions are sufficient conditions for the original potential. They are not necessary, and therefore might be throwing away part of the parameter space. However, it still gives us a very good sense of the possibilities within the Type-Z 3HDM.

 \subsubsection{Unitarity}
 In order to determine the tree-level unitarity constraints, we use
 the algorithm presented in \cite{Bento:2017eti}. As described there,
 we have to impose that the eigenvalues of the scattering S-matrix of
 two scalars into two scalars have an upper bound (the unitarity
 limit). As these arise exclusively from the quartic part of the
 potential, the eigenvalues obtained for a $\Z3$ symmetric potential
 in Section 4.4 of \cite{Bento:2017eti} can also be used for the
 potential with quadratic soft-breaking terms, \eq{Z3potential}. The
 conversion between the notation of the algorithm and the potential
 chosen, \eq{Z3quartic}, is as follows: 
\ba
r_1 \to \lambda_1 \; ,  \qquad r_2 \to \lambda_2 \; ,  \qquad r_3 \to \lambda_3 ,\\[8pt]
r_4 \to \lambda_4/2 \; ,  \qquad r_5 \to \lambda_5/2 \; ,  \qquad r_6 \to \lambda_6/2 ,\\[8pt]
r_7 \to \lambda_7/2 \; ,  \qquad r_8 \to \lambda_8/2 \; ,  \qquad r_9 \to \lambda_9/2 ,\\[8pt]
c_4 \to \lambda_{10}/2 \; ,  \qquad c_{12} \to \lambda_{11}/2 \; ,  \qquad c_{11} \to \lambda_{12}/2 .
\ea
Denoting by $\Lambda_i$ the eigenvalues of the relevant scattering matrices,
we have 21 $\Lambda$'s to calculate for each set of physical
parameters randomly generated, and the condition to impose is that 
\begin{equation}
     |\Lambda_i| \leq 8\pi\,\,,\quad i=1,..,21\, .
\end{equation}

\subsubsection{Oblique parameters $STU$}

In order to discuss the effect of the $S,T,U$ parameters,
we use the results in \cite{Grimus:2007if}.
To apply the relevant expressions,
we write the matrices $U$ and $V$ used in \cite{Grimus:2007if}
with the notation choices that we made when obtaining the mass
eigenstates in section~\ref{subsec:scalar}.
We start with the $3\times 6$ matrix $V$ defined as
  \begin{equation}
      \begin{pmatrix}x_1+\,i\,z_1 \\ x_2+\,i\,z_2 \\ x_3+\,i\,z_3  \end{pmatrix} = V \begin{pmatrix}G^0 \\ h_1 \\ h_2 \\ h_3 \\ A_1 \\ A_2 \end{pmatrix} ,
  \end{equation}
and find, by comparison with \eqs{CPevenDiag}{CPoddDiag}, that $V$ is 
 \begin{equation}
     V= \begin{pmatrix}[1.5] i \textbf{P}^T_{11} & \textbf{R}^T_{11}& \textbf{R}^T_{12}&\textbf{R}^T_{13} &  i \textbf{P}^T_{12} &  i \textbf{P}^T_{13} \\ i \textbf{P}^T_{21} & \textbf{R}^T_{21}& \textbf{R}^T_{22}&\textbf{R}^T_{23} &  i \textbf{P}^T_{22} &  i \textbf{P}^T_{23}\\i \textbf{P}^T_{31} & \textbf{R}^T_{31}& \textbf{R}^T_{32}&\textbf{R}^T_{33} &  i \textbf{P}^T_{32} &  i \textbf{P}^T_{33}  \end{pmatrix} .
 \end{equation}
  
The $3\times 3$ matrix U defined as
\begin{equation}
\begin{pmatrix} w_1^\dagger \\ w_2^\dagger \\ w_3^\dagger \end{pmatrix}
=U \begin{pmatrix} G^\dagger \\ H_1^+ \\ H_2^+  \end{pmatrix},
\label{chargedtransfU}
\end{equation}
gives us the correspondence $U=\textbf{Q}^T $ from \eq{ChargedDiag}.
  
Having applied the expressions for $S, T, U $,
the constraints implemented on $S$ and $T$ follow
Fig.~4 of Ref.~\cite{Baak:2014ora}, at $95\%$ confidence level.
For $U$, we fix the allowed interval to be
\begin{equation}
    U=0.03\pm0.10 .
\end{equation}

\subsection{Theoretical Constraints 2}

As we want to explore the range of low $\tan\beta_1$ and $\tan\beta_2$
we should avoid that the Yukawa couplings become non-perturbative. We
have, in our model
\begin{align}
      Y_t=&\frac{m_t\sqrt{2}}{v}\,
      \frac{\sqrt{1+\tan\beta_2^2}}{\tan\beta_2}\, ,\\[+2mm]
      Y_\tau=&\frac{m_\tau\sqrt{2}}{v}\,\sqrt{1+\tan\beta_1^2}\,
      \sqrt{1+\tan\beta_2^2}
      \, ,\\[+2mm] 
      Y_b=&\frac{m_b\sqrt{2}}{v}\,
      \frac{\sqrt{1+\tan\beta_1^2}\, \sqrt{1+\tan\beta_2^2}}{\tan\beta_1} \, .
\end{align}
We require
\begin{equation}
  \label{eq:1}
  \frac{Y^2}{4\pi} < 1 \quad \Rightarrow \quad Y< \sqrt{4\pi}
\end{equation}

\subsection{$\Delta M_{b,s}$ Constraints}

We see from Fig.~1 of Ref.~\cite{Chakraborti:2021bpy} that
the constraints coming from $\Delta M_{b,s}$ tend to
exclude very low values on $\tan\beta$. Thus, we take
\begin{equation}
  \label{eq:2}
  \log_{10}(\tan\beta_{1,2}) >-0.5\quad \Rightarrow \quad
  \tan\beta_{1,2} > 10^{-0.5} = 0.31623\, .
\end{equation}

\subsection{LHC Constraints}
For comparison with experiment,  we consider only the contributions of the
lowest non-vanishing order in perturbation theory. 
The decays that require one-loop calculations are those of neutral scalars
into two photons ($h_j\to\gamma\gamma$),
one Z and one photon ($h_j\to Z\gamma$),
and two gluons ($h_j\to gg$).
The final formulas for the first two widths are given in
Ref.~\cite{Fontes:2014xva},
only having to adapt the particles and their couplings to our case.
The formula for the width $h_j\to\gamma\gamma$ reads,
\begin{equation}
\Gamma(h_j\to \gamma\gamma)=\frac{G_F\alpha^2m_h^3}{128\sqrt{2}\pi^3}
(|X_F^{\gamma\gamma}+X_W^{\gamma\gamma}+X_H^{\gamma\gamma}|^2),
\label{eq:gaga}
\end{equation}
where, noticing that for scalars the $Y$ terms in \cite{Fontes:2014xva} vanish,
\ba
X_F^{\gamma\gamma}
&=&
-\sum_f N_c^f2a_j^fQ_f^2\tau_f[1+(1-\tau_f)f(\tau_f)],
\\[8pt]
X_{W}^{\gamma \gamma}
&=&
C_{j}\left[2+3 \tau_{W}+3 \tau_{W}\left(2-\tau_{W}\right)
f\left(\tau_{W}\right)\right],
\label{need_Cj}
\\[8pt]
X_{H}^{\gamma \gamma}
&=&
-\sum_{k=1}^{2} \frac{\lambda_{h_j H_k^+ H_k^-} v^{2}}{2 m_{H_{k}^{\pm}}^{2}}
\tau_{j k}^{\pm}
\left[1-\tau_{j k}^{\pm} f\left(\tau_{j k}^{\pm}\right)
\right].
\label{XHformula}
\ea
We used
\begin{equation}
\tau_f = 4m_f^2/m_{h_j}^2\, ,
\ \ \ 
\tau_{j k}^{\pm} = 4m_{H_k^\pm}^2/m_{h_j}^2\, ,
\end{equation}
where $m_f$ ($m_{H_k^\pm}$) is the mass of the relevant particle
in the loop, while $m_{h_j}$ is the mass of the decaying
Higgs boson. The function $f(\tau)$ is defined in
the Higgs Hunter's Guide \cite{Gunion:1989we},
\begin{equation}
f(\tau)=\left\{\begin{array}{ll}
{\left[\sin ^{-1}(\sqrt{1 / \tau})\right]^{2},} & \text { if } \tau \geq 1 \\
-\frac{1}{4}\left[\ln \left(\frac{1+\sqrt{1-\tau}}{1-\sqrt{1-\tau}}\right)-i \pi\right]^{2}, & \text { if } \tau<1
\end{array}\right. ,
\end{equation}
and the couplings $C_{j}$ and $\lambda_{h_j H_k^+ H_k^-}$
for this model are written in the appendix. They were derived with the
help of the 
software \texttt{FeynMaster}\cite{Fontes:2019wqh,Fontes:2021iue}, that
uses \texttt{QGRAF}\cite{Nogueira:1991ex},
\texttt{FeynRules}\cite{Christensen:2008py,Alloul:2013bka} and
\texttt{FeynCalc}\cite{Mertig:1990an,Shtabovenko:2016sxi}  in an
integrated way.

The decay into gluons can be obtained from the expression for the $\gamma\gamma$ decay,
\begin{equation}
    \Gamma(h_j\to gg)=\frac{G_F\alpha_S^2m_h^3}{64\sqrt{2}\pi^3}(|X_F^{gg}|^2) ,
\end{equation}
where
\begin{equation}
    X_F^{gg}=-\sum_q2a_j^q\tau_q[1+(1-\tau_q)f(\tau_q)] ,
\end{equation}
and the sum runs only over quarks q. 

For the 125GeV scalar, the coupling  modifiers,
are  calculated  directly  from  the  random  angles 
generated and constrained to be within $2\sigma$ of the most
recent ATLAS fit results, \cite[Table 10]{Aad:2019mbh}. 
Having chosen a specific production and decay channel,
the collider event rates can be conveniently  described by the cross section ratios $\mu_{if}^h$,
\beq
\mu_{if}^h=\left(\frac{\sigma_i^{\text{3HDM}}(pp\to h) }{\sigma_i^{\text{SM}}(pp\to h)}\right)\left(\frac{\text{BR}^{\text{3HDM}}(h\to f)}{\text{BR}^{\text{SM}}(h\to f)}\right).
\eeq
Starting from the collision of two protons, the relevant production
mechanisms include: gluon fusion (ggH), vector boson fusion (VBF),
associated production with a vector boson (VH, V = W or Z), and
associated production with a pair of top quarks (ttH). The SM cross
section for the gluon fusion process is calculated using HIGLU
\cite{Spira:1995mt}, and for the other production mechanisms we use
the results of Ref.~\cite{deFlorian:2016spz}.  Each of the 3HDM
processes is obtained by rescaling the SM cross sections by the
relevant relative couplings. As for the decay channels, we calculated
the branching rations for final states $f=\,W\,W,\, Z\,Z,\,
b\,\overline{b}, \gamma\,\gamma$ and $\tau^+\tau^-$. Finally, we
require that the $\mu_{if}^h$ for each individual initial state
$\times$ final state combination is consistent, within twice the total
uncertainty, with the best-fit results presented in the most recent
study of data collected at $\sqrt{s}=13\,\text{TeV}$ with the ATLAS
experiment \cite[Figure 5]{Aad:2019mbh}. 

For the heavier neutral and charged scalars,
we use \texttt{HiggsBounds-5.9.1} in Ref.~\cite{Bechtle:2020pkv},
where a list of all the relevant experimental analyses
can be found.
For the decays allowing for off-shell bosons,
we use the method explained in \cite{Romao:1998sr}.
We also consider the constraints coming from $b\to s
\gamma$, as we explain in sections~\ref{sect:xsgamma} and \ref{sec:Fig2}.

\section{\label{sec:hgaga}Decays of $h_{125}$ in the $\Z3$ 3HDM}

In this section, we use the scan ranges defined in Eq.~(\ref{scanparameters}),
pass them through all theoretical and experimental constraints,
and we study the impact on the decays of the 125GeV Higgs $h_1=h_{125}$
found at LHC.

\begin{figure}[H]
\centering
\includegraphics[width = 0.5\textwidth]{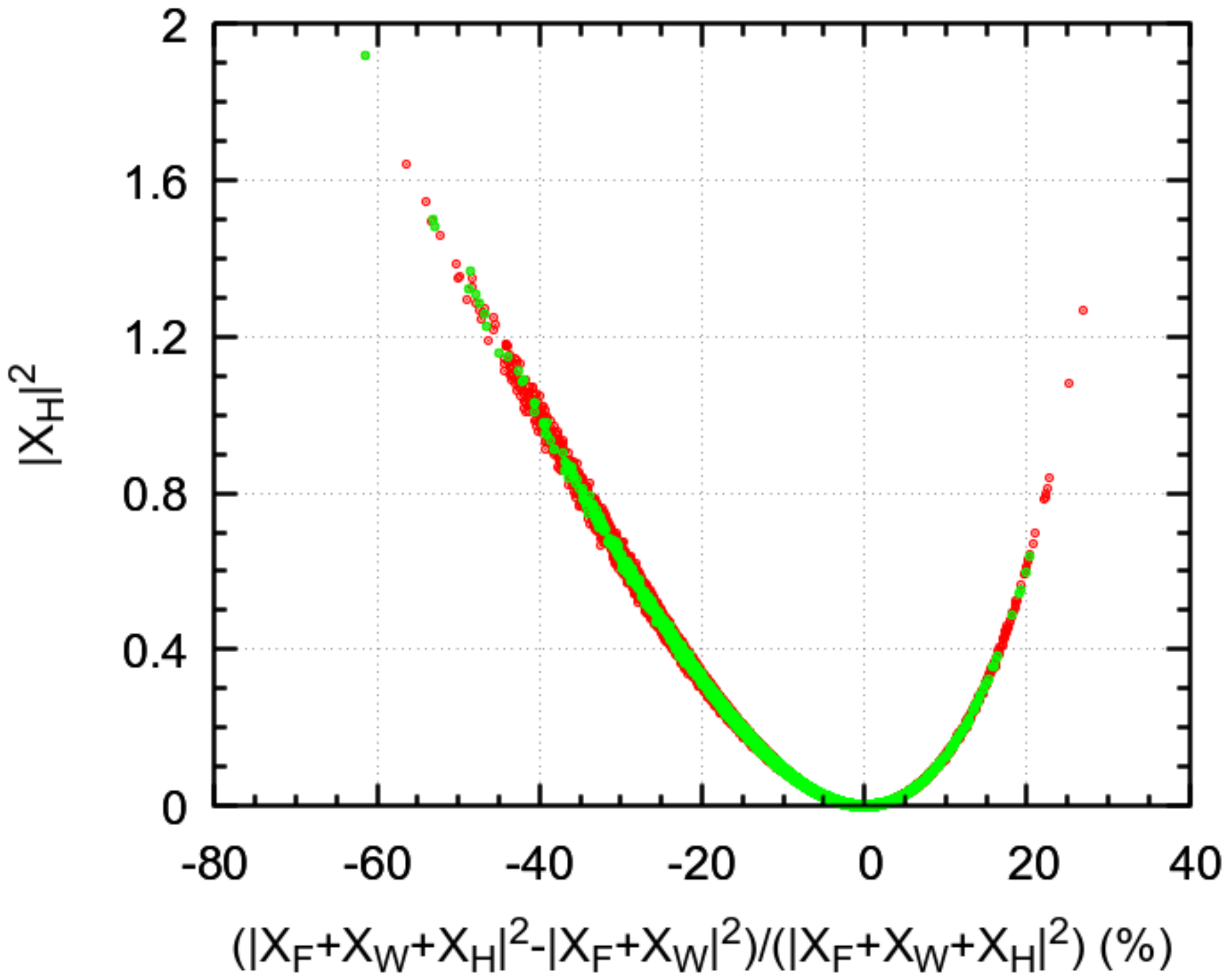}
\caption{\label{HtoGaGa}Effect of the charged Higgs on the
$h_{125}\to\gamma\gamma$ decay, with the definitions of \eq{eq:gaga}.
The green points passed all constraints including HB5,
while the red points did not pass HB5.}
\end{figure}

The contribution from the two charged scalars to the $h_{125}\to\gamma\gamma$ decay process is shown in Fig.~\ref{HtoGaGa}. There are two interesting regimes.
To the left (right) of the vertical line at coordinate zero,
the two charged Higgs conspire to decrease (increase) the branching
ratio into $\gamma\gamma$.
Most of the points are on the left and correspond
to a significant reduction of the decay width.
However, there are indeed points on the right,
which allow for an increase which could be up by 20\%. 
We have also confirmed the existence of allowed results where
the destructive interference between the two charged Higgs
leads to a null $X_H$, occurring when the signs of the
couplings $\lambda_{h_jH_1^+H_1^-}$ and $\lambda_{h_jH_2^+H_2^-}$
are opposite in \eq{XHformula}.
This means that, barring other constraints,
the charged Higgs masses could be relatively light without
contradicting the observed $h_{125}\to\gamma\gamma$,
as long as their contributions to this decay canceled,
as they may.
\noindent
The points of Fig.~\ref{HtoGaGa} where $|X_H|^2$ is large,
for which the charged Higgs provide a considerable contribution
to the overall $h_{125}\to\gamma\gamma$ decay rate (the latter, still
within current bounds) is only obtained for very fine tuned
points in parameter space with some charged Higgs mass
below 200GeV.
As we will see in Figs.~\ref{fig:6}-\ref{fig:7} below,
this is a very constrained (fine tuned) region.

\begin{figure}[H]
\centering
\includegraphics[width = 0.5\textwidth]{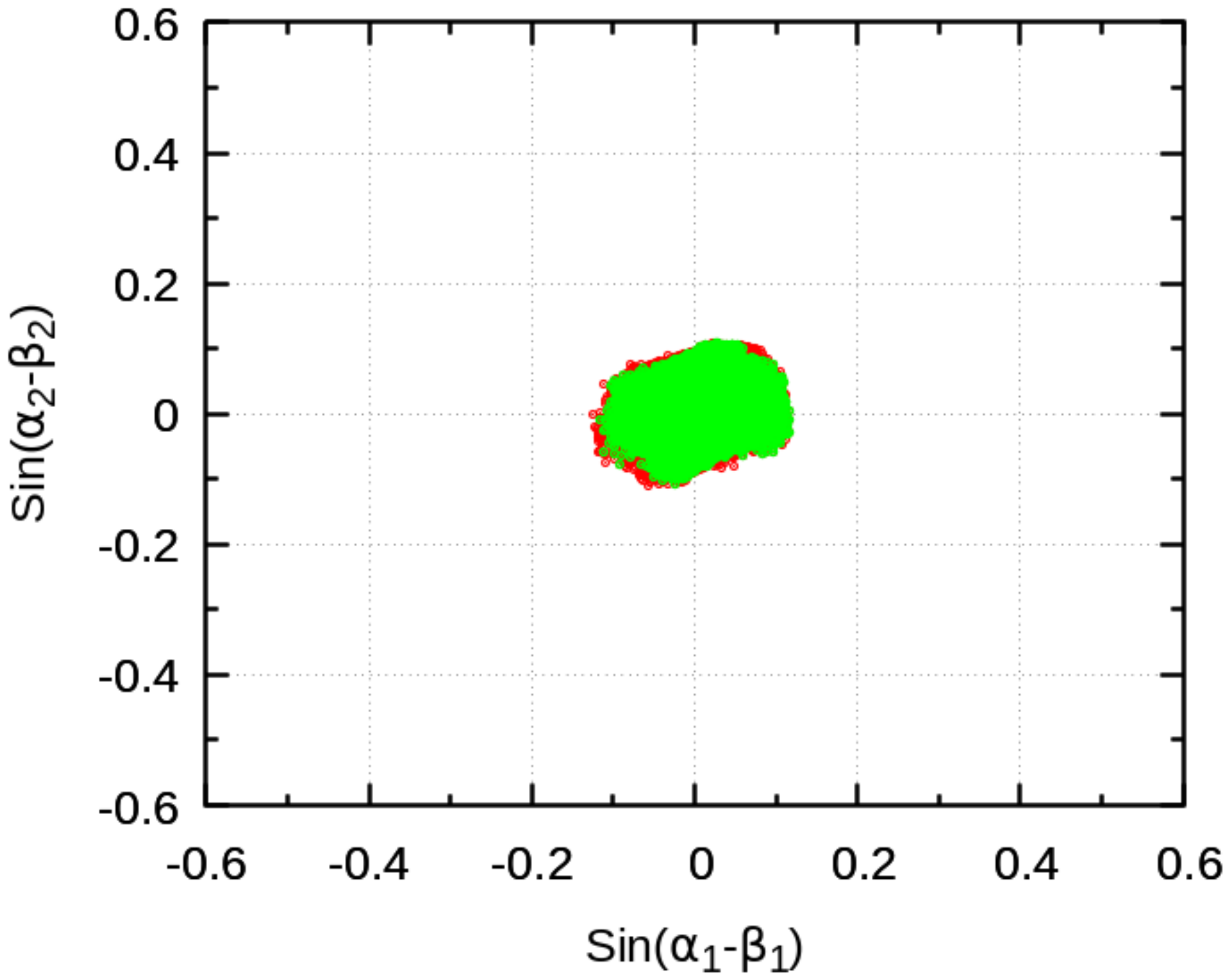}
\caption{Results of the simulation in the $\sin{(\alpha_2-\beta_2)}-\sin{(\alpha_1-\beta_1)}$ plane. The green points passed all constraints including HB5, while the red points did not pass HB5. \label{plota1a2}}
\end{figure}

The set of points that are consistent with all the bounds is now
plotted in the $\sin{(\alpha_2-\beta_2)}-\sin{(\alpha_1-\beta_1)}$
plane as shown in Fig.~\ref{plota1a2}. Comparing with the plot in the
same plane shown in \cite[Fig.1]{Das:2019yad}, it can be seen that the
use of more recent experimental data for the simulated results leads
to us being closer to the alignment limit, defined by
$\alpha_1=\beta_1$ and $\alpha_2=\beta_2$. 

However, as we will illustrate below, points in parameter space
close to the alignment limit exhibit physical properties which differ
significantly from the exact alignment limit.

To study the allowed regions for the cross section ratios $\mu_{if}^h$, we follow \cite{Fontes:2014xva,Barroso:2012wz} and calculate each $\mu_{if}^h$ using all production channels. Our set of points is then  shown in Figs.~\ref{fig:gaga-zz} - \ref{fig:gaga-Zga}. Similar to the complex 2HDM analyzed by Fontes, Rom\~{a}o and Silva
in \cite{Fontes:2014xva}, there is a strong correlation between $\mu_{Z\gamma}$ and $\mu_{\gamma\gamma}$ in our Type-Z model, as shown in Fig.~\ref{fig:gaga-Zga}. 
\begin{table}[H]
	\begin{minipage}{0.45\linewidth}
     \centering
    \includegraphics[width = 0.95\textwidth]{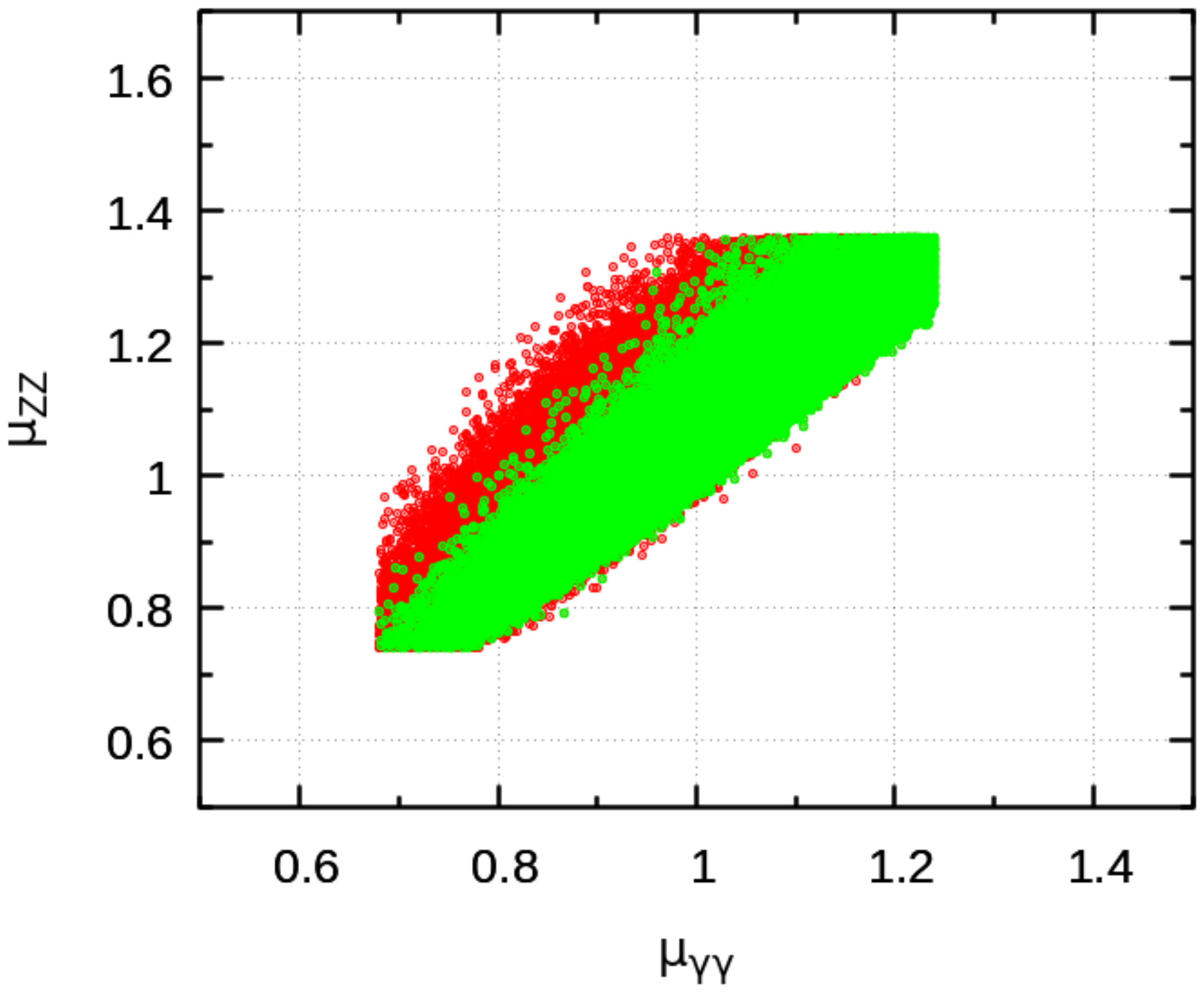}
    \captionof{figure}{Results in the $\mu_{ZZ}-\mu_{\gamma\gamma}$ plane for the gluon fusion production channel. The green points passed all constraints
    including HB5, while the red points do not include HB5.}
    \label{fig:gaga-zz}
	\end{minipage}\hfill
    \begin{minipage}{0.45\linewidth}
     \centering
    \includegraphics[width = 0.95\textwidth]{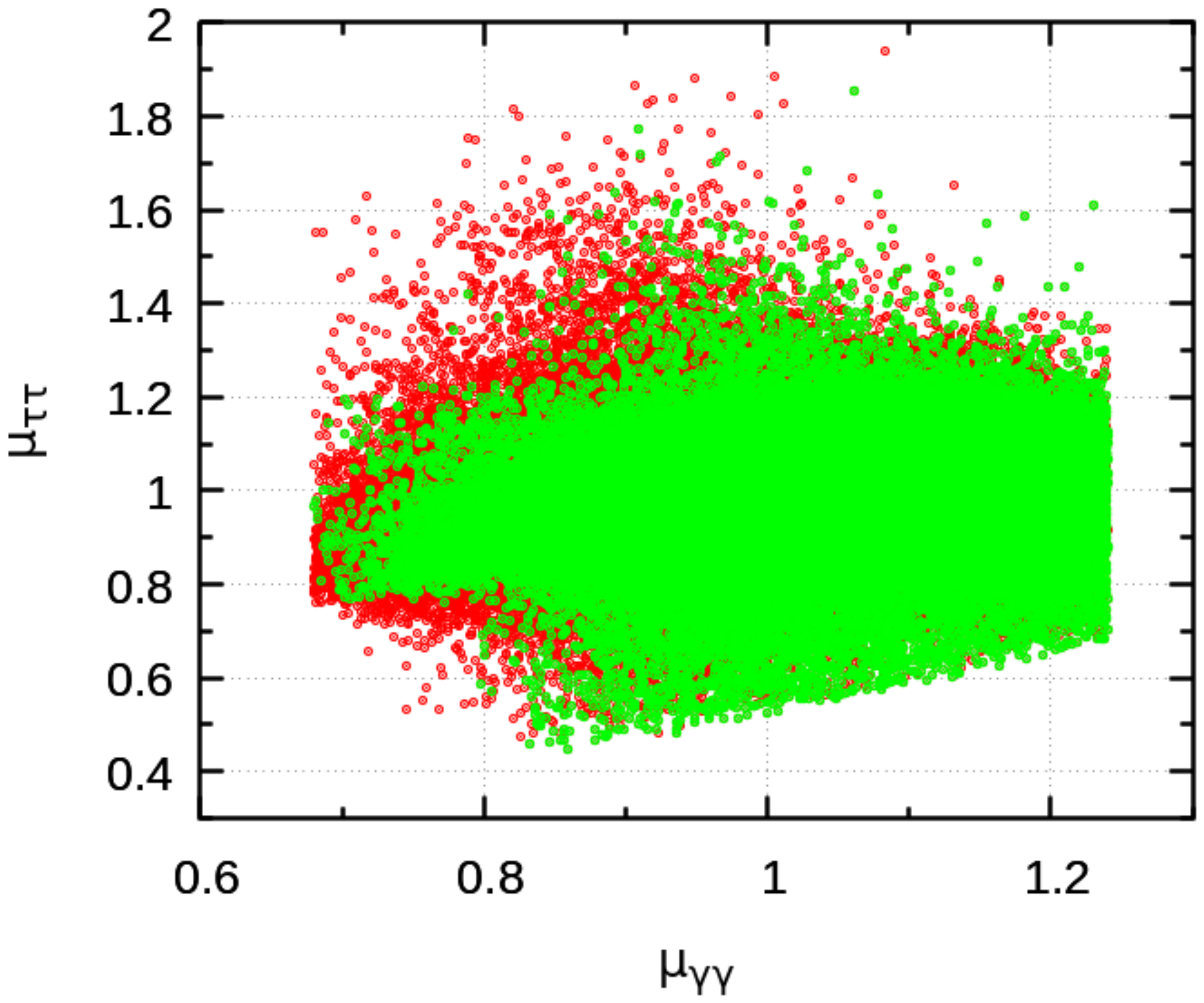}
    \captionof{figure}{Results in the $\mu_{\tau\tau}-\mu_{\gamma\gamma}$ plane for all production channels. The green points passed all constraints
        including HB5, while the red points do not include HB5.}
    \label{fig:gaga-tautau}
	\end{minipage}
\end{table}
\begin{table}[H]
	\begin{minipage}{0.45\linewidth}
     \centering
    \includegraphics[width = 0.95\textwidth]{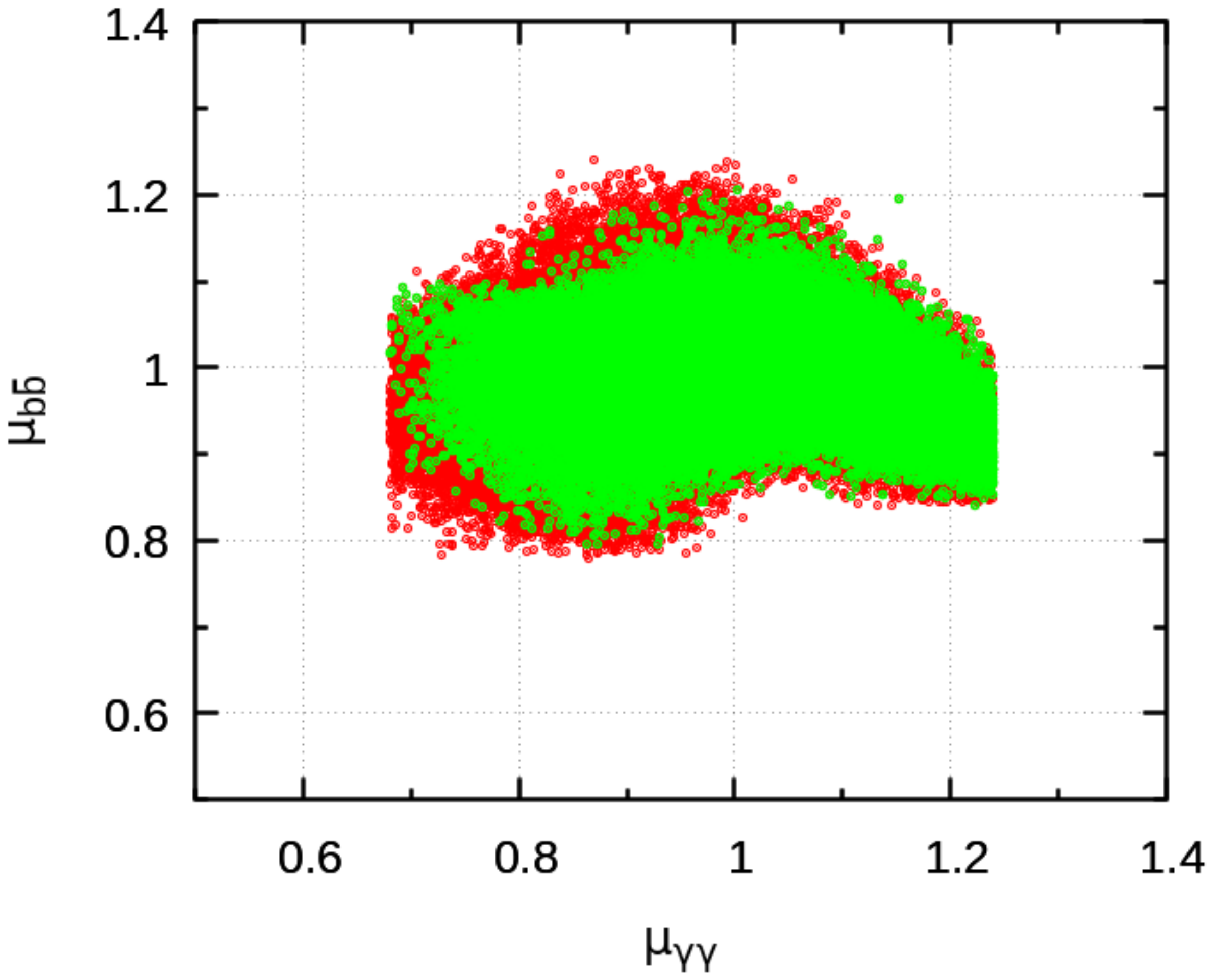}
    \captionof{figure}{Results in the $\mu_{b\overline{b}}-\mu_{\gamma\gamma}$ plane for the gluon fusion production channel. The green points passed all constraints
        including HB5, while the red points do not include HB5.}
    \label{fig:gaga-bb}
	\end{minipage}\hfill
    \begin{minipage}{0.45\linewidth}
     \centering
    \includegraphics[width = 0.95\textwidth]{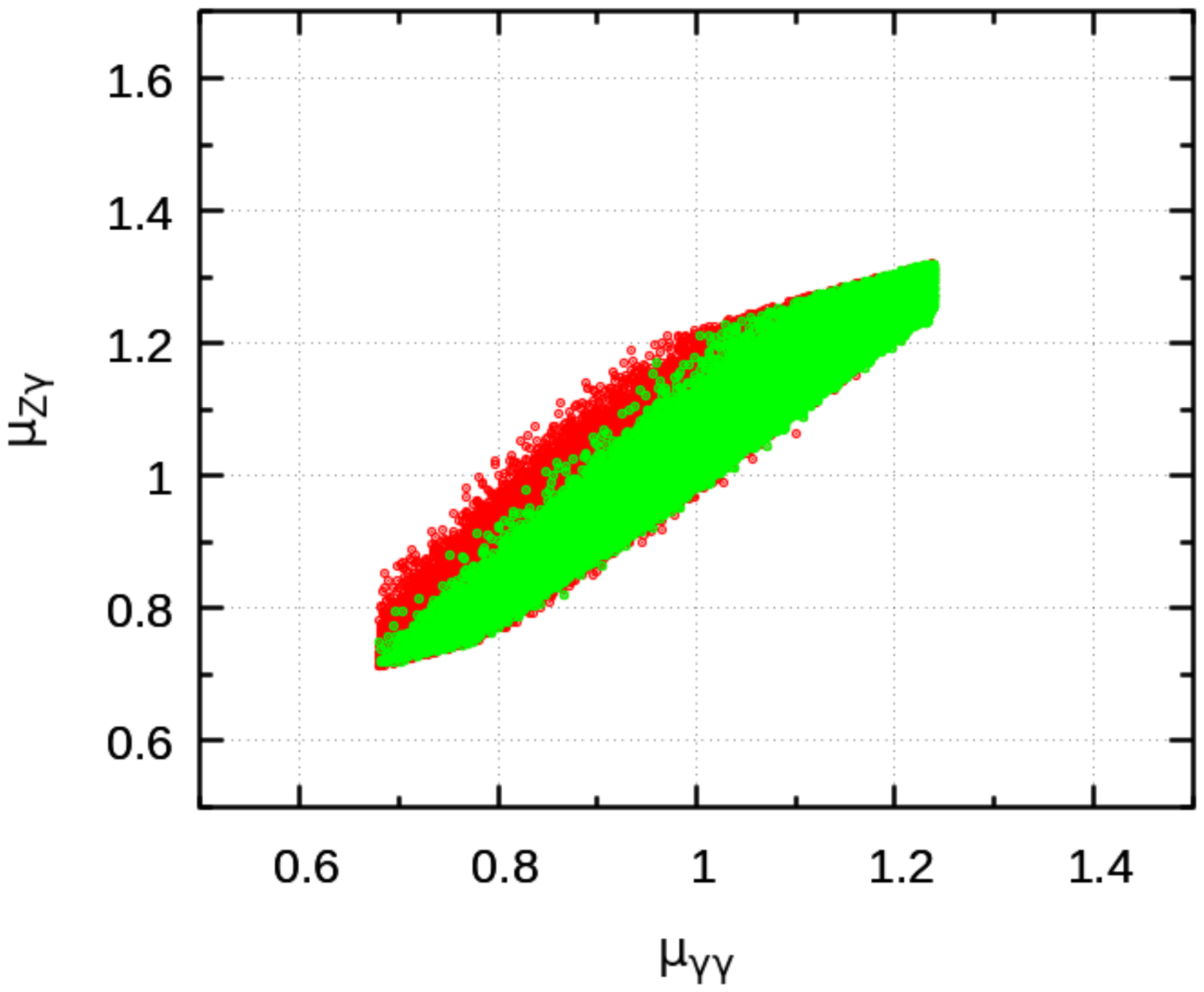}
    \captionof{figure}{Results in the $\mu_{Z\gamma}-\mu_{\gamma\gamma}$ plane for the gluon fusion production channel. The green points passed all constraints
        including HB5, while the red points do not include HB5.}
    \label{fig:gaga-Zga}
	\end{minipage}
\end{table}
Such a correlation is also visible between $\mu_{ZZ}$ and $\mu_{\gamma\gamma}$
in Fig.~\ref{fig:gaga-zz}.
It is less apparent in correlations with $\tau^+ \tau^-$
and $b \bar{b}$,
as shown in Figs.~\ref{fig:gaga-tautau} and \ref{fig:gaga-bb}.

\section{\label{sect:xsgamma}Calculation of the BR($B\to X_s \gamma$)}

\subsection{Introduction}

It is well known that the experimental bounds on $B\to X_s \gamma$
place stringent restrictions on the parameter space of models with charged
scalars~\cite{Borzumati:1998tg,Borzumati:1998nx,
Misiak:2017bgg,Misiak:2018cec,Akeroyd:2020nfj}.
Most notably,
there is a bound on the mass of the only charged Higgs boson
present in the Type-II 2HDM which, at 95\% CL
(2$\sigma$), is according to \cite{Misiak:2017bgg} 
\begin{equation}
  \label{eq:16}
  m_{H^+}> 580\, \text{GeV}\, .
\end{equation}
The exact value for this bound depends on both the theoretical 
approximations \cite{Bernlochner:2020jlt} and the experimental errors.
The experimental average gives \cite{Amhis:2019ckw}
\begin{equation}
  \label{eq:25}
  \text{BR}^{\rm exp}(B\to X_s \gamma) = (3.32 \pm 0.15) \times 10^{-4}\, ,
\end{equation}
while the NNLO calculation within the SM yields
\cite{Misiak:2020vlo,Akeroyd:2020nfj}
\begin{equation}
  \label{eq:23}
  \text{BR}^{\rm SM}(B\to X_s \gamma) = (3.40 \pm 0.17) \times 10^{-4}\, ,
\end{equation}
with an error of about 5\%.

As explained below,
we will take an error of 2.5\% around the central value of the
calculation and, following \cite{Akeroyd:2020nfj}, 
we consider 99\% CL (3$\sigma$) for the experimental error:
\begin{equation}
  \label{eq:17}
  2.87 \times 10^{-4} < \text{BR}(B\to X_s \gamma) < 3.77 \times 10^{-4}\, .
\end{equation}

\subsection{The calculation}

We follow closely the calculation by Borzumati and Greub in
Ref.~\cite{Borzumati:1998tg}.
There, the new contributions from the charged Higgs bosons are encoded
in the Wilson coefficients,
\begin{subequations}   
\begin{align}
  \label{eq:18}
  C^{0,{\rm eff}}_7(\mu_W) =&
  C^{0,{\rm eff}}_{7,\rm SM}(\mu_W) +|Y|^2 C^{0,{\rm eff}}_{7,\rm YY}(\mu_W)
  +(X Y^*)C^{0,{\rm eff}}_{7,\rm XY}(\mu_W)\, , \\[+2mm]
  C^{0,{\rm eff}}_8(\mu_W) =&
  C^{0,{\rm eff}}_{8,\rm SM}(\mu_W) +|Y|^2 C^{0,{\rm eff}}_{8,\rm YY}(\mu_W)
  +(X Y^*)C^{0,{\rm eff}}_{8,\rm XY}(\mu_W)\, , \\[+2mm]
  C^{1,{\rm eff}}_4(\mu_W) =&
  E_0(x) + \frac{2}{3}
  \log\left(\frac{\mu_W^2}{M_W^2}\right) +|Y|^2 E_H(y)\, ,\\[+2mm]  
  C^{1,{\rm eff}}_7(\mu_W) =&
  C^{1,{\rm eff}}_{7,\rm SM}(\mu_W) +|Y|^2 C^{1,{\rm eff}}_{7,\rm YY}(\mu_W)
  +(X Y^*)C^{1,{\rm eff}}_{7,\rm XY}(\mu_W)\, , \\[+2mm]
  C^{1,{\rm eff}}_8(\mu_W) =&
  C^{1,{\rm eff}}_{8,\rm SM}(\mu_W) +|Y|^2 C^{1,{\rm eff}}_{8,\rm YY}(\mu_W)
  +(X Y^*)C^{1,{\rm eff}}_{8,\rm XY}(\mu_W) \, ,
\end{align}
\end{subequations}   
where we are using the notation in Ref.~\cite{Borzumati:1998tg}
which should be consulted for the definitions and also for the procedure
used in evolving the coefficients to the scale $\mu_b=m_b$.
The dependence on the charged Higgs mass
appears because the functions $C^{0,{\rm eff}}_{i,\rm YY},
C^{0,{\rm eff}}_{i,\rm XY}, C^{1,{\rm eff}}_{i,\rm YY}$, and
$C^{1,{\rm eff}}_{i,\rm XY}$ depend on $y=m_t^2/m_{H^+}^2$,
while the SM coefficients depend on $x= m_t^2/M_W^2$.

For models with multiple charged Higgs there is one contribution
(and one parameter $y_k$) for each particle.
A model with two charged Higgs is discussed in \cite{Akeroyd:2020nfj,Logan:2020mdz},
with interesting earlier work highlighting the possible cancellation
between the two charged Higgs contributions appearing in
Refs.~\cite{Hewett:1994bd,Akeroyd:2016ssd}.
We obtain, for example,
\begin{align}
  \label{eq:19}
  C^{1,{\rm eff}}_7(\mu_W) =&
  C^{1,{\rm eff}}_{7,\rm SM}(\mu_W) +|Y_1|^2 C^{1,{\rm eff}}_{7,\rm  YY}(\mu_W,y_1)
  +|Y_2|^2 C^{1,{\rm eff}}_{7,\rm YY}(\mu_W,y_2)\nonumber\\
& +(X_1 Y_1^*)C^{1,{\rm eff}}_{7,\rm XY}(\mu_W,y_1) 
  +(X_2 Y_2^*)C^{1,{\rm eff}}_{7,\rm XY}(\mu_W,y_2) \, ,
\end{align}
where we wrote explicitly the dependence on the
charged Higgs masses,
\begin{equation}
  \label{eq:20}
  y_1= \frac{m_t^2}{m_{H_1^+}^2}, \quad
    y_2= \frac{m_t^2}{m_{H_2^+}^2}\, ,
\end{equation}
and used
\begin{equation}
X_1 = -\frac{\textbf{Q}_{22}}{\cos\beta_2 \sin\beta_1}, \quad
      Y_1 =  \frac{\textbf{Q}_{23}}{\sin\beta_2}, \quad X_2 = -\frac{\textbf{Q}_{32}}{\cos\beta_2\,\sin\beta_1}, \quad Y_2 =  \frac{\textbf{Q}_{33}}{\sin\beta_2}.
\end{equation}

We took the input parameters from Ref.~\cite{Borzumati:1998tg} except for
$\alpha_s(M_Z),m_t,M_Z,M_W$, that were updated to the most recent
values of the Particle Data Group
\cite{Zyla:2020zbs}:\footnote{If we use exclusively the input values of
Ref.~\cite{Borzumati:1998tg},
we reproduce their SM results.
We are extremely grateful to C. Greub for discussions and for providing
us with the original code used in \cite{Borzumati:1998tg},
utilized to cross check our independent calculations.}
\begin{subequations}   
\begin{align}
  \label{eq:21}
 & \alpha_s(M_Z)= 0.1179\pm 0.0010, && m_t=172.76 \pm 0.3 \, \text{GeV},\\
  &m_c/m_b= 0.29 \pm 0.02,  && m_b-m_c = 3.39 \pm 0.04\, \text{GeV},\\
  &\alpha_{em}^{-1}=137.036, && |V_{ts}^*V_{tb}/V_{cb}|^2=0.95\pm
  0.03,\\
  &\text{BR}_{SL}= 0.1049\pm 0.0046\, .&& &&
\end{align}
\end{subequations}   

\section{\label{sec:Fig2}Impact of $b \rightarrow s \gamma$ 
on the $\Z3$ 3HDM parameter space}

We find that much of the parameter space considered in
Ref.~\cite{Chakraborti:2021bpy} is forbidden. This
is most apparent by considering their Fig.~2,
which we turn to next.

\subsection{Only $b\to s \gamma$}

\begin{figure}[H]
    \centering
    \begin{tabular}{cc}
      \includegraphics[width=0.48\textwidth]{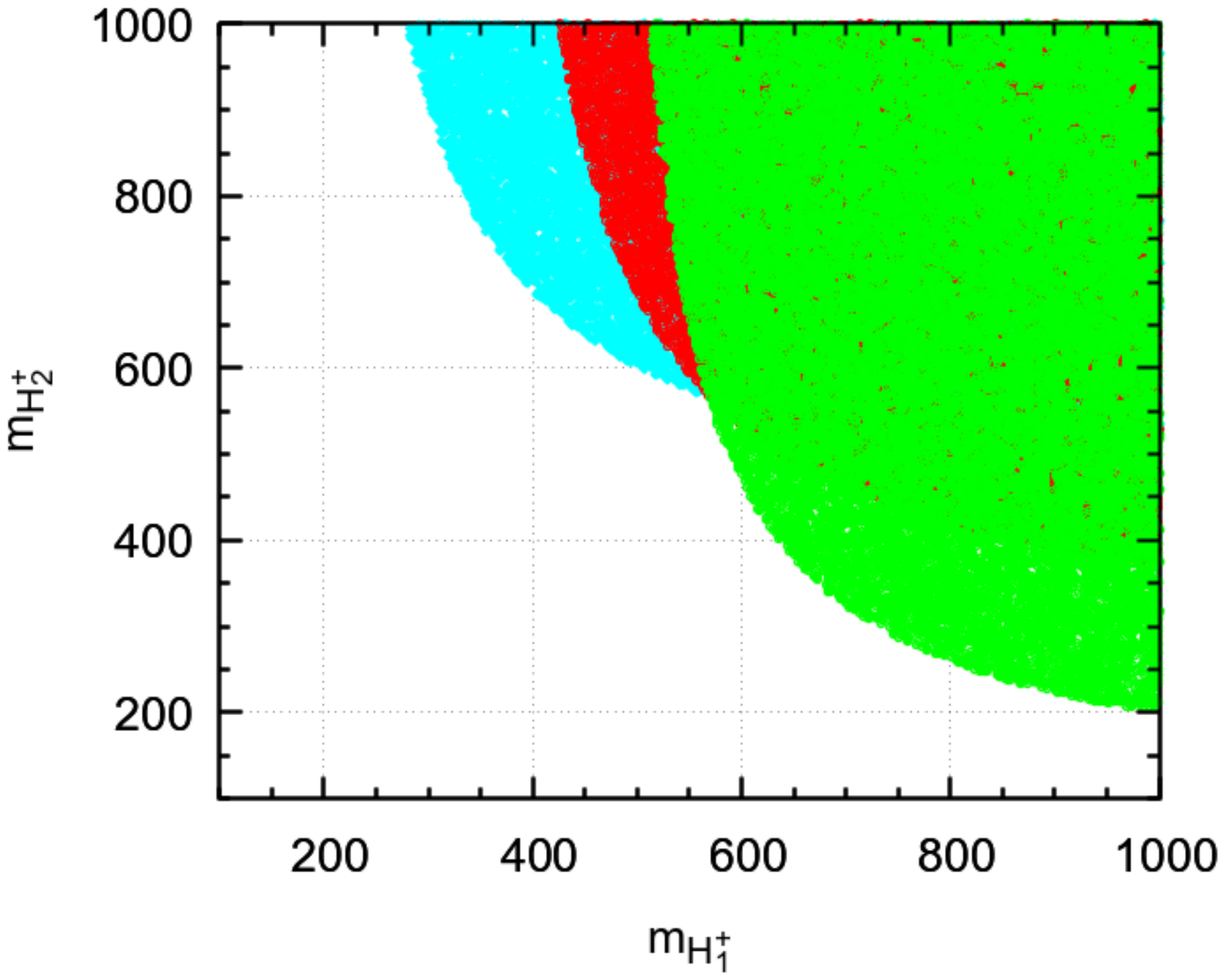}
      &
        \includegraphics[width=0.48\textwidth]{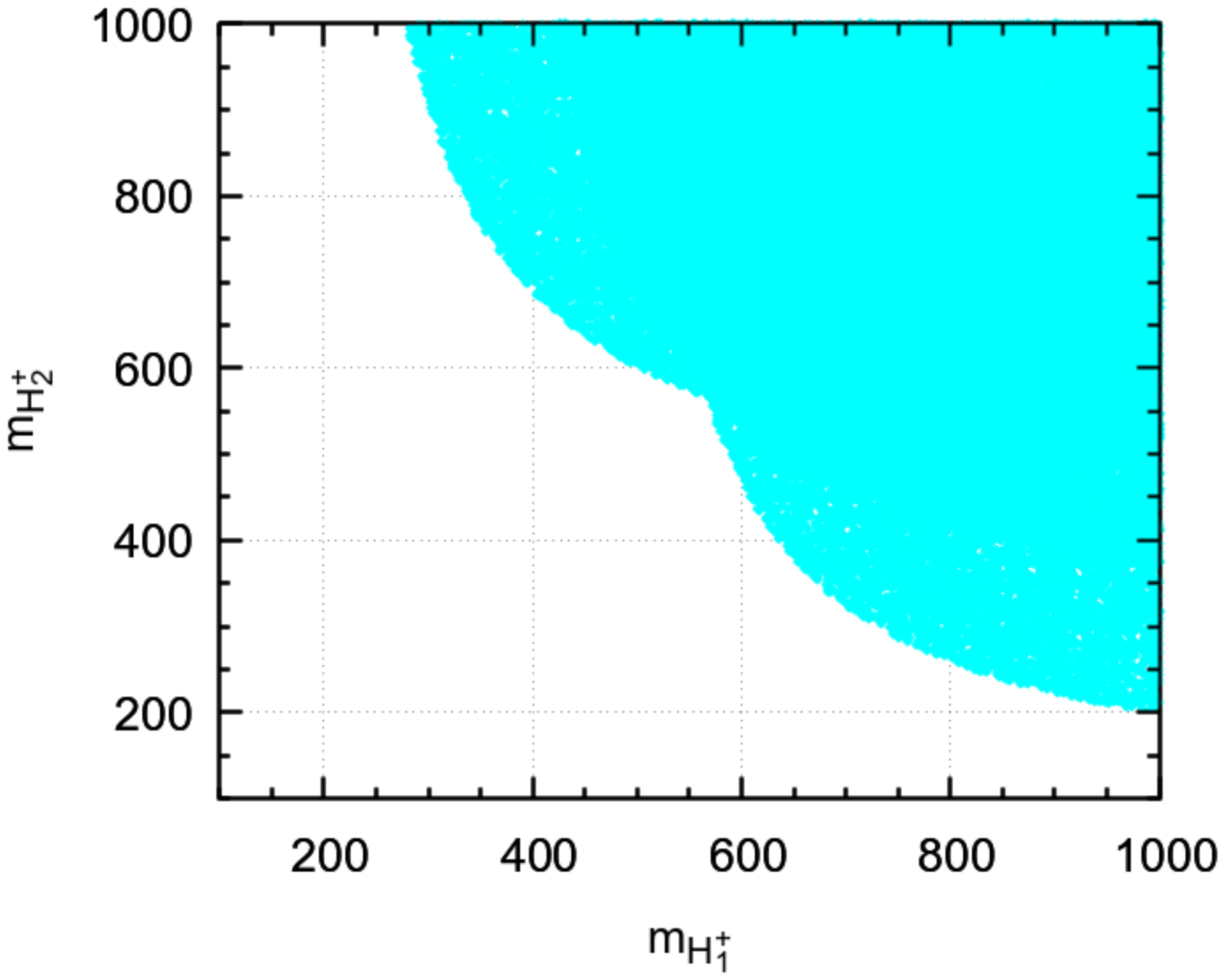}
    \end{tabular}
    
    \caption{Comparison with Fig.~2 of
      Ref.~\cite{Chakraborti:2021bpy}. On the left panel we have
      $\gamma_2= \frac{\pi}{6}$ (cyan),  $\gamma_2= \frac{\pi}{4}$
      (red), $\gamma_2= \frac{\pi}{3}$ (green). On the right panel the 3 regions
    are superimposed.}
    \label{fig:1}
  \end{figure}

On Fig.~2 of Ref.~\cite{Chakraborti:2021bpy} the parameters are fixed as
\begin{equation}
  \label{eq:3}
  \tan\beta_1=10,\quad\tan\beta_2=2,\quad
  \gamma_2=\frac{\pi}{6},\frac{\pi}{4},\frac{\pi}{3} \, ,
\end{equation}
while imposing
\begin{equation}
  \label{eq:4}
  m_{h_2} = m_{A_1}=m_{H_1^+},\quad m_{h_3} =
  m_{A_2}=m_{H_2^+},\quad
  \alpha_1=\beta_1, \alpha_2=\beta_2,\gamma_1=\gamma_2=-\alpha_3 \, .
\end{equation}
Applying only the $b\to s \gamma$ cut we reproduce their Fig.~2
in our Fig.~\ref{fig:1}.
\noindent
Fig.~\ref{fig:1} passes all theoretical constraints,
even including unitarity and BFB.

\subsection{The effect of other constraints}

From the previous plots, the conclusion that we can have
one of the charged Higgs relatively light if the other is sufficiently
heavy seems correct. However we now show that for this choice of
parameters this is not the case. With the choice of
Eqs.~(\ref{eq:3})-(\ref{eq:4},
the bounds from the decays of the 125 GeV Higgs are simply satisfied.
However the same is not true for current bounds on heavier scalars.
Indeed, every single point in Fig.~\ref{fig:1} is excluded by HB5; not
a single point remains.
This will be explained in detail in the following section.

\subsection{Enlarging the region of good points}

\begin{figure}[H]
  \centering
  \begin{tabular}{cc}
      \includegraphics[width=0.48\textwidth]{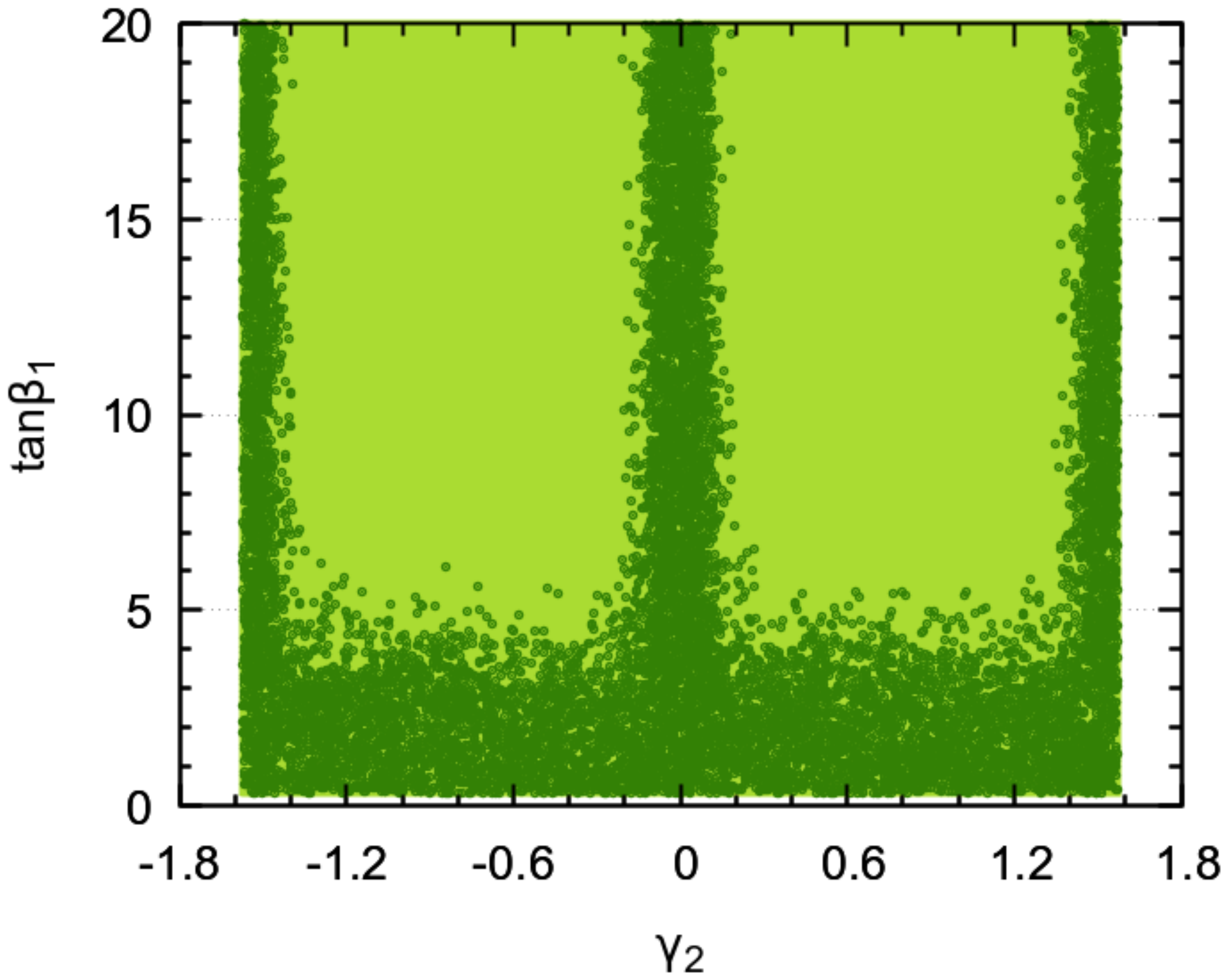}
      &
      \includegraphics[width=0.48\textwidth]{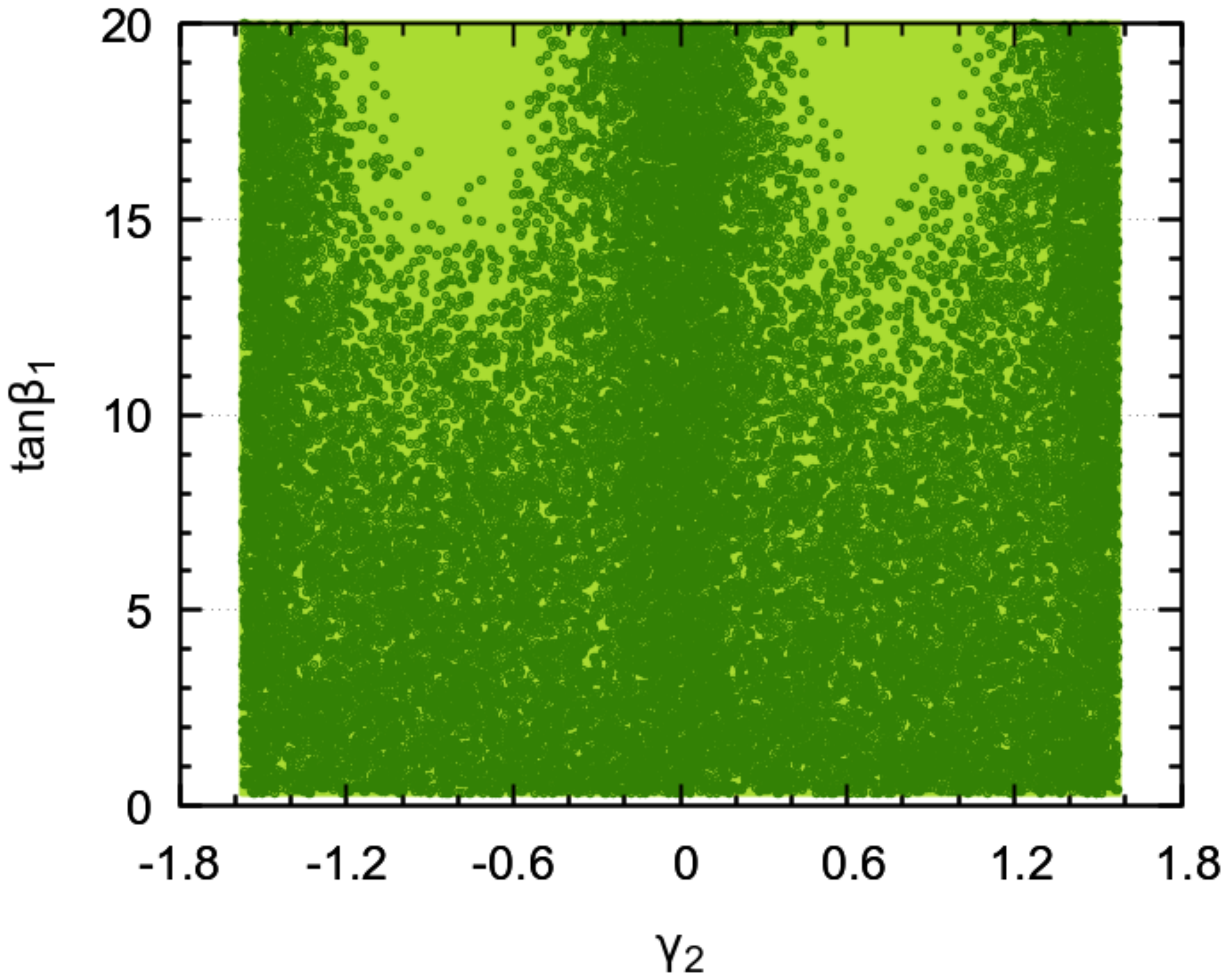}
   \end{tabular}
  \caption{\label{fig:new}Enlarging good points, taking $\gamma_2 \in
      [-\pi/2,\pi/2]$ and varying $\beta_1$, all other conditions in
      Eqs.~(\ref{eq:3})-(\ref{eq:4}) were
      kept. The dark green points passed all constraints including HB5,
      while the light green points did not pass HB5. Left panel: All points
      passing \texttt{HiggsBounds-5.9.1}. Right panel: All points
      passing \texttt{HiggsBounds-5.7.1}.}
\end{figure}

We discovered that the situation described in the previous section is
a consequence of the small range chosen for $\gamma_2$. To illustrate this, we
kept the other conditions in Eqs.~(\ref{eq:3})-(\ref{eq:4}),
but allowed for
\begin{equation}
  \label{eq:5}
  \gamma_2 \in [-\pi/2,\pi/2] \, ,
\end{equation}
and (for Fig.~\ref{fig:new}) also varied $\tan{\beta_1}$.
The points which survive \texttt{HiggsBounds-5.9.1} are shown in dark green on
the left panel of Fig.~\ref{fig:new}.
\noindent
The allowed points for $\tan{\beta_1}=10$ are concentrated
around $\gamma_2 = 0, \pm\pi/2$,
excluding $\gamma_2 = \pi/6,\pi/4,\pi/3$.
Taking the interval in Eq.~(\ref{eq:5}) one can indeed find
regions of good points.\footnote{This it true regardless of whether or not
we vary $\beta_1$, as long as we enlarge the region of $\gamma_2$.}

It is interesting to compare with what happens with the previous
version of \texttt{HiggsBounds-5.7.1}, shown on
the right panel of Fig.~\ref{fig:new}.
For that case there are many points allowed for all values of $\gamma_2$,
even for $\tan{\beta_1}=10$.
We have found that this is due to the recent bounds on
$h_{2,3} \rightarrow \tau^+ \tau^-$ decay in
Ref.~\cite{ATLAS:2020zms},
included in \texttt{HiggsBounds-5.9.1} but not in
\texttt{HiggsBounds-5.7.1},
which used the previous
bounds~\cite{CMS:2015mca,CMS:2017epy}.\footnote{In
Ref.~\cite{Chakraborti:2021bpy} the strong constraints from neutral
scalar decays 
into $\tau\tau$ still seemed to allow points with the choices in
Eqs.~(\ref{eq:3})-(\ref{eq:4}).
}

\begin{figure}[H]
  \centering
  \begin{tabular}{cc}
    \includegraphics[width=0.48\textwidth]{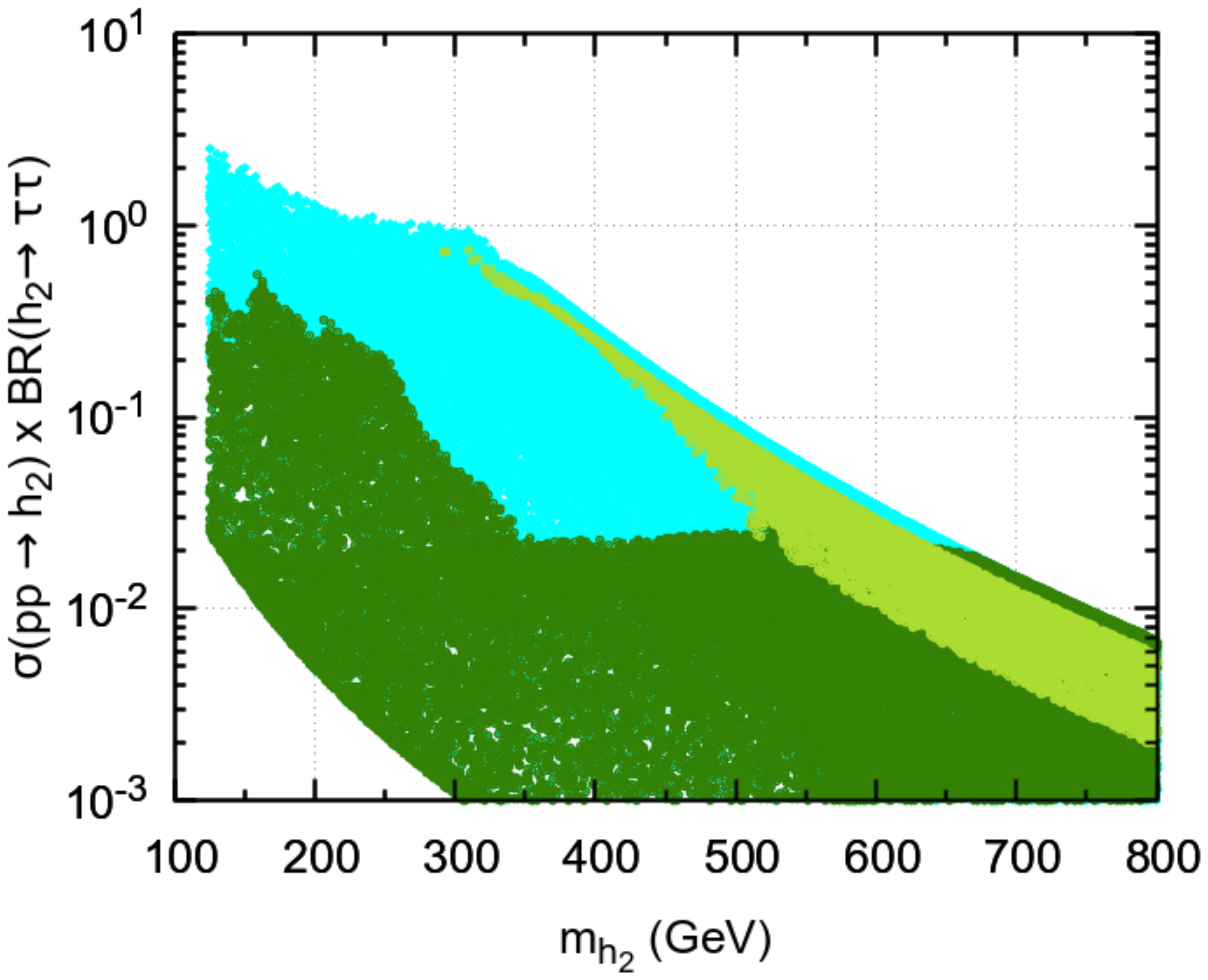}
    &
    \includegraphics[width=0.48\textwidth]{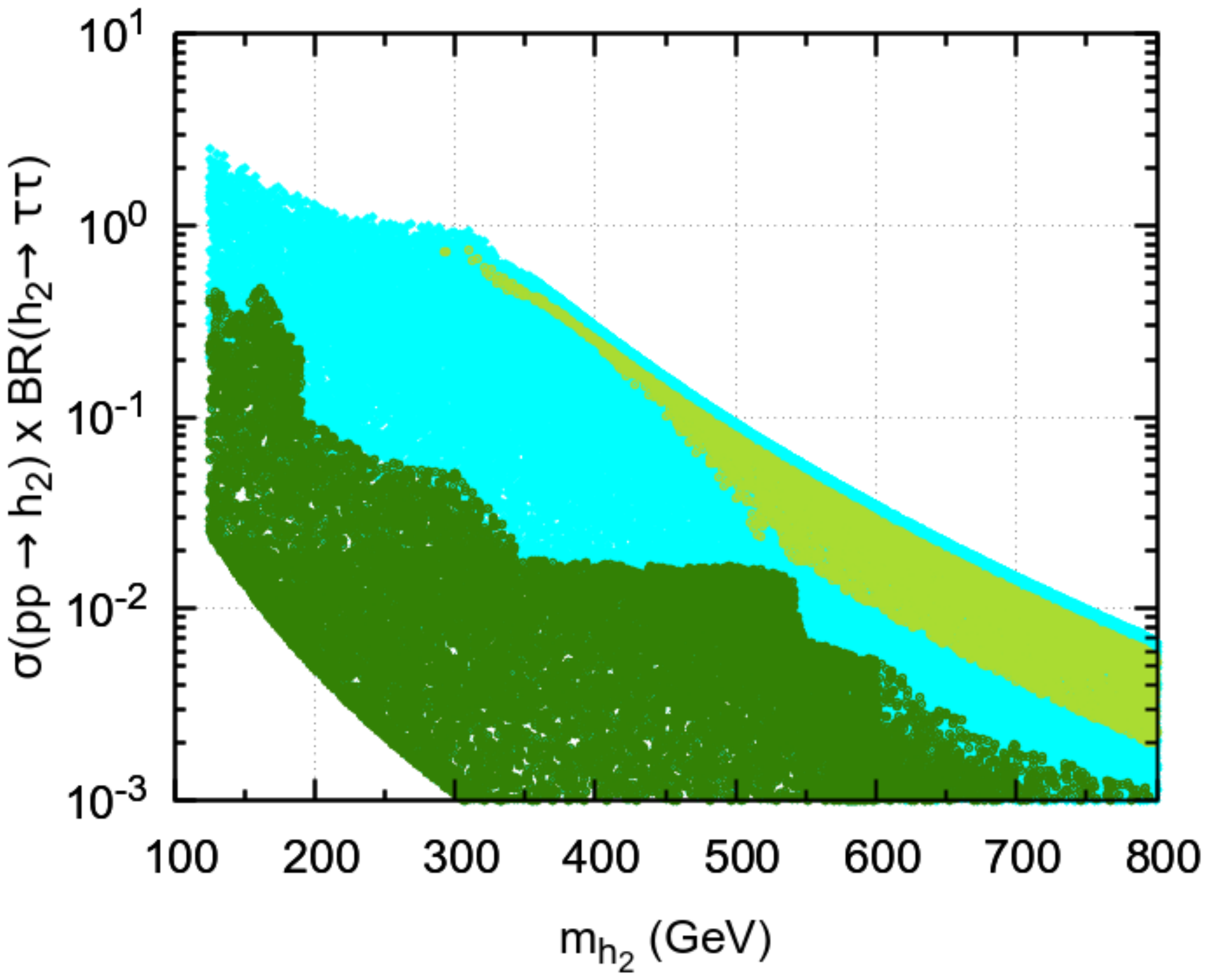}
  \end{tabular}
  \caption{\label{fig:sigBR}
    Left Panel: $\sigma(pp\to h_2) \times \text{BR}(h_2 \to \tau\tau)$
    as function of the $m_{h_2}$. Parameters are as in Eq.~(\ref{eq:4}),
    except that $\gamma_2 \in [-\pi/2,\pi/2]$. Points in
    cyan are points that pass all constraints before
    \texttt{HiggsBounds}
    and in dark
    green after \texttt{HiggsBounds-5.7.1}. In light green are the points in the
    interval $\gamma_2 
    \in [\pi/6,\pi/3]$. Right Panel: the same but for
    \texttt{HiggsBounds-5.9.1}}  
\end{figure}

To better illuminate this point,
we show  $\sigma(pp\to h_2) \times \text{BR}(h_2 \to \tau\tau)$
versus $m_{h_2}$ in Fig.~\ref{fig:sigBR}.
In this figure, the parameters are as in Eqs.~(\ref{eq:3})-(\ref{eq:4}, 
except that $\gamma_2 \in [-\pi/2,\pi/2]$.
Points in cyan are points that pass all constraints before
\texttt{HiggsBounds}.
In light green are the points in the restricted
interval $\gamma_2  \in [\pi/6,\pi/3]$.
In the left panel points in  
dark green  are those who survided after
\texttt{HiggsBounds-5.7.1}.
In the right panel we have
the same situation but now we used \texttt{HiggsBounds-5.9.1}. We see that
there were good points in the restricted interval $\gamma_2  \in
[\pi/6,\pi/3]$ in the left panel, but they disappeared with the
newer version \texttt{HiggsBounds-5.9.1}. We have confirmed that
similar plots can be obtained for $h_3$.
This is a good point to stress again the role that the LHC is having in
constraining models with new scalar physics.
One sees the strong impact that the updated LHC results have
in constraining the $\Z3$ 3HDM.
This highlights the importance that the new LHC run will have
in constraining the parameter space of extended scalar sectors.

\begin{figure}[H]
\centering
\includegraphics[width = 0.5\textwidth]{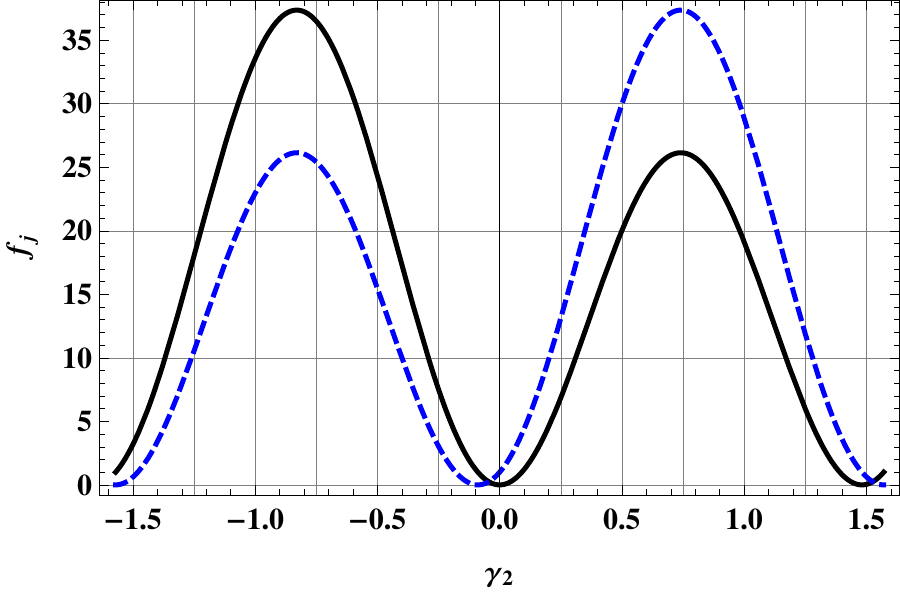}
\caption{\label{fig:fj}Graphic of the functions $f_j$ defined
in Eq.~(\ref{eq:7}) for varying $\gamma_2=-\alpha_3$, with
$\tan{\beta_1} = 10$ and 
$\tan{\beta_2} = 2$.
Function $f_2$ ($f_3$) in black/solid (blue/dashed) line.}
\end{figure}

To better understand the behaviour of 
$\sigma(pp\to h_i) \times \text{BR}(h_i \to \tau\tau)$ ($i=2,3$) we can make
the simplified assumption\footnote{We are neglecting the dependence of
the cross section on the mass.} that this product is proportional to
\begin{equation}
  \label{eq:7}
  \sigma(pp\to h_i) \times \text{BR}(h_i \to \tau\tau)
  \propto
  g_{h_i \tau \tau}^2\, g_{h_i t t}^2 \equiv f_i\ ,
\end{equation}
where we are assuming that the production occurs mainly
via gluon fusion with the top quark in the loop.
Now, using the assumptions of Eq.~(\ref{eq:4})
in Eq.~(\ref{coeffNeutralFerm}), we have
\begin{align}
  \label{eq:8}
  &g_{h_2 \tau\tau} = - \frac{c_{\alpha_3} t_{\beta_1}}{c_{\alpha_2}}
  - s_{\alpha_3} t_{\beta_2}
= -\frac{t_{\beta_1}}{c_{\beta_2}}\, c_{\gamma_2} +t_{\beta_2}
s_{\gamma_2}\, ,
  && g_{h_2 tt} = \frac{c_{\alpha_2} s_{\alpha_3}}{s_{\beta_2}} = -
  \frac{1}{t_{\beta_2}}\, s_{\gamma_2}
  \ ,\nonumber\\[+2mm]
  &g_{h_3 \tau\tau} = 
-c_{\alpha_3} t_{\beta_2}+
\frac{s_{\alpha_3}t_{\beta_1}}{c_{\alpha_2}}
=- t_{\beta_2} c_{\gamma_2} - \frac{t_{\beta_1}}{c_{\beta_2}}
s_{\gamma_2} \, ,
&& g_{h_3 tt} = \frac{c_{\alpha_2} c_{\alpha_3}}{s_{\beta_2}}
= \frac{1}{t_{\beta_2}}\, c_{\gamma_2}
  \ ,
\end{align}
where, for Fig.~\ref{fig:sigBR}, $\beta_1,\beta_2$ are fixed and
$\gamma_2 \in [-\pi/2,\pi/2]$.
Fig.~\ref{fig:fj} shows the functions in Eq.~(\ref{eq:7}) --
$f_2$ for $h_2$ and $f_3$ for $h_3$ -- for $\tan{\beta_1} = 10$ and
$\tan{\beta_2} = 2$ as in Eq.~\eqref{eq:3},
but keeping $\gamma_2$ free.
\noindent
We see that these functions are largest precisely in the approximate interval
$\pm \gamma_2 \in [\pi/6, \pi/3]$. This explains why these points are the first
to be excluded by the bounds on
$\sigma(pp\to h_2) \times \text{BR}(h_2 \to \tau\tau)$,
and why, going outside such bounds, some points can be
preserved.\footnote{Of course, we have ignored in this simple reasoning the
dependence on $m_{h_i}$, which has been taken into account appropriately
in our scans and HB5 limits.}

\subsection{The effect of $\tan\beta$'s}

\begin{figure}[H]
  \centering
  \begin{tabular}{cc}
      \includegraphics[width=0.48\textwidth]{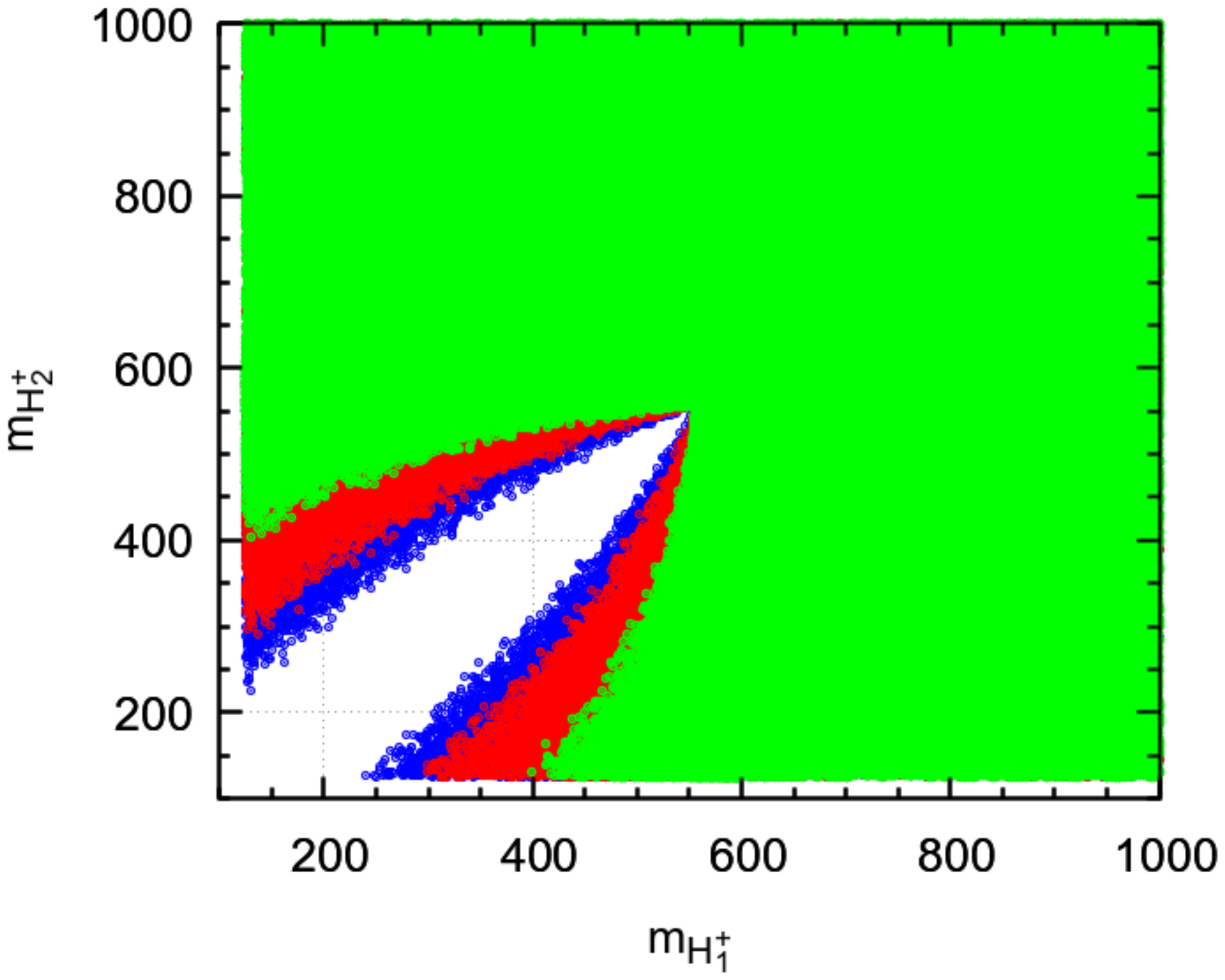}
      &
      \includegraphics[width=0.48\textwidth]{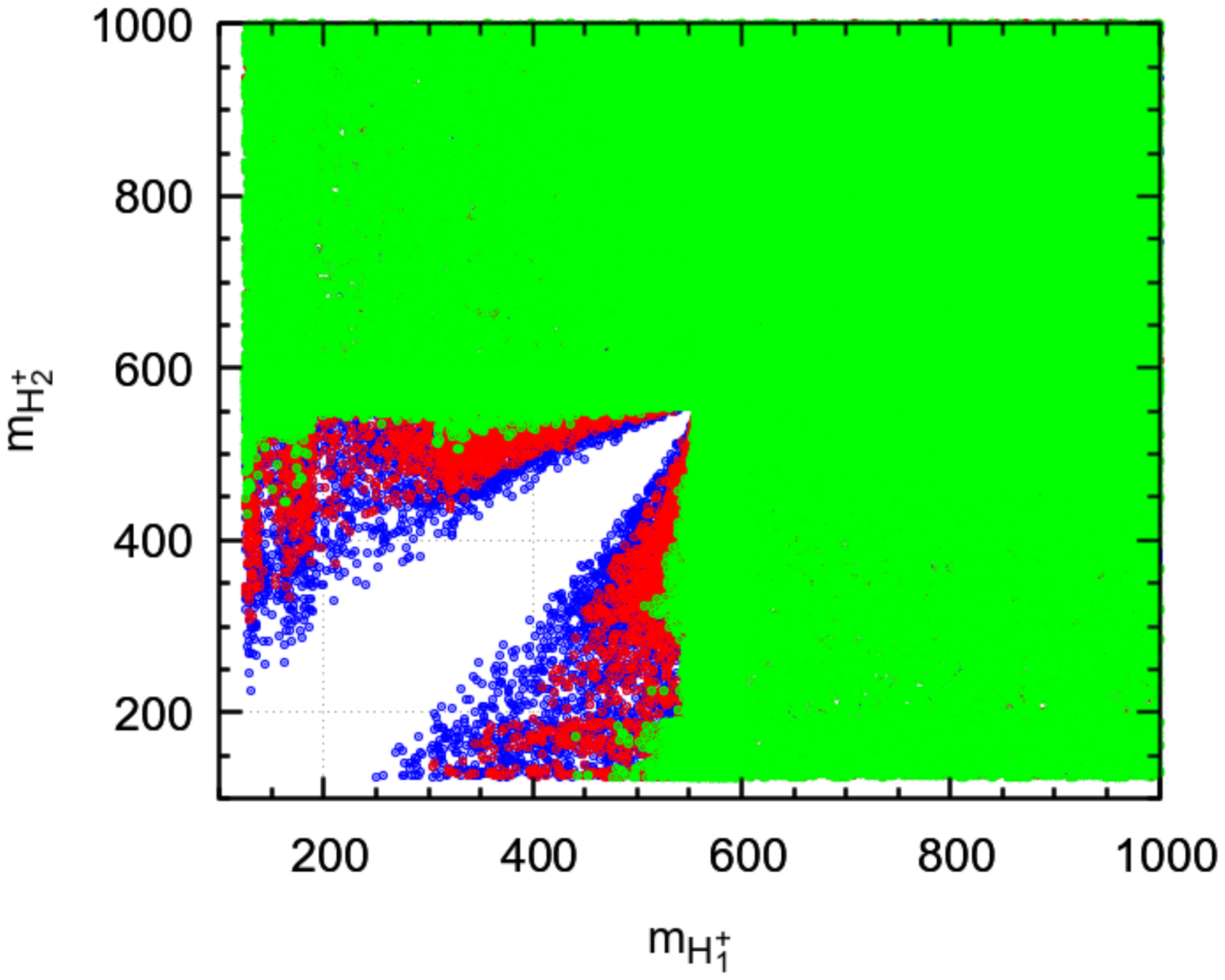}
   \end{tabular}
  \caption{All points satisfy Eq.~(\ref{eq:4}). Left panel: All points
    passed all constraints except for 
    HB5.
    The blue
    points satisfy Eq.~(\ref{eq:6}). The red points are for
    $\tan\beta_{1,2}>0.5$ and the green points
    are for $\tan\beta_{1,2}>1$. Right panel: same color code as in
    the left panel but only showing points surviving after requiring
    HB5.}
  \label{fig:5}
\end{figure}

In the last section we saw that while maintaining the main features of
Eqs.~(\ref{eq:3})-(\ref{eq:4}, but enlarging the range of variation of
$\gamma_2$, 
we could find points allowed by all current experimental constraints.
Here we exploit the variation of both
$\tan\beta$'s in the range
\begin{equation}
  \label{eq:6}
  \tan\beta_{1,2} \in [10^{-0.5},10] \, ,
\end{equation}
subject to the condition of perturbativity of the Yukawa couplings in
Eq.~(\ref{eq:1}). The result is shown in Fig.~\ref{fig:5}. 
\noindent
We see that by varying the range of $\tan\beta$'s we can have smaller
masses for the charged Higgs bosons. For $\tan\beta< 1$ it is even possible
to have both charged Higgs with masses below 400 GeV.

\section{\label{sec:beyondalign}Going beyond exact alignment}

We have performed a completely uniform scan and found out that very
few points survived and those were not too far away from the alignment
condition of Eq.~(\ref{eq:4}). So another strategy can be to scan
points that differ from the perfect alignment of Eq.~(\ref{eq:4}) by
1\% or 10\%.

\begin{figure}[H]
  \centering
  \begin{tabular}{cc}
      \includegraphics[width=0.48\textwidth]{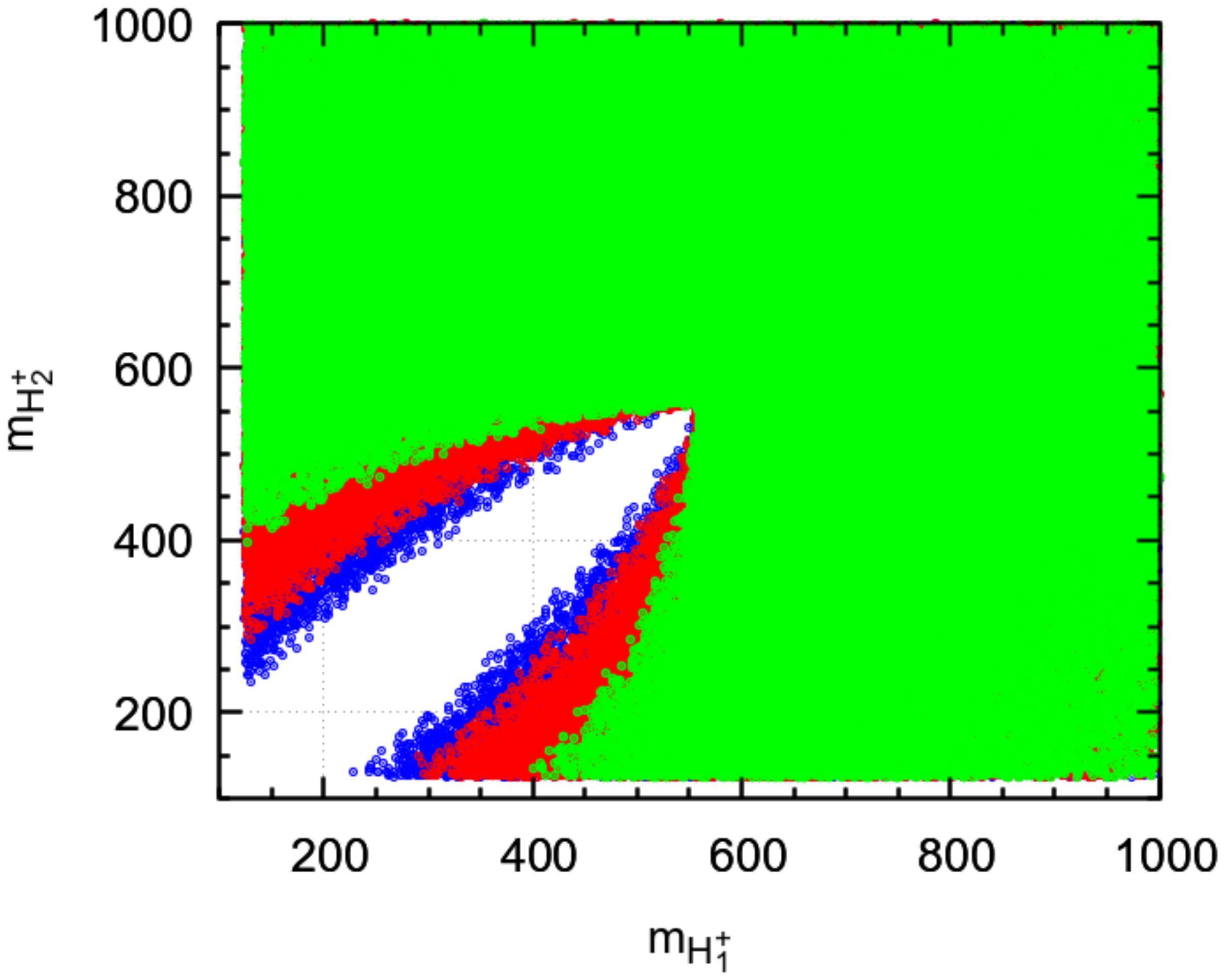}
      &
      \includegraphics[width=0.48\textwidth]{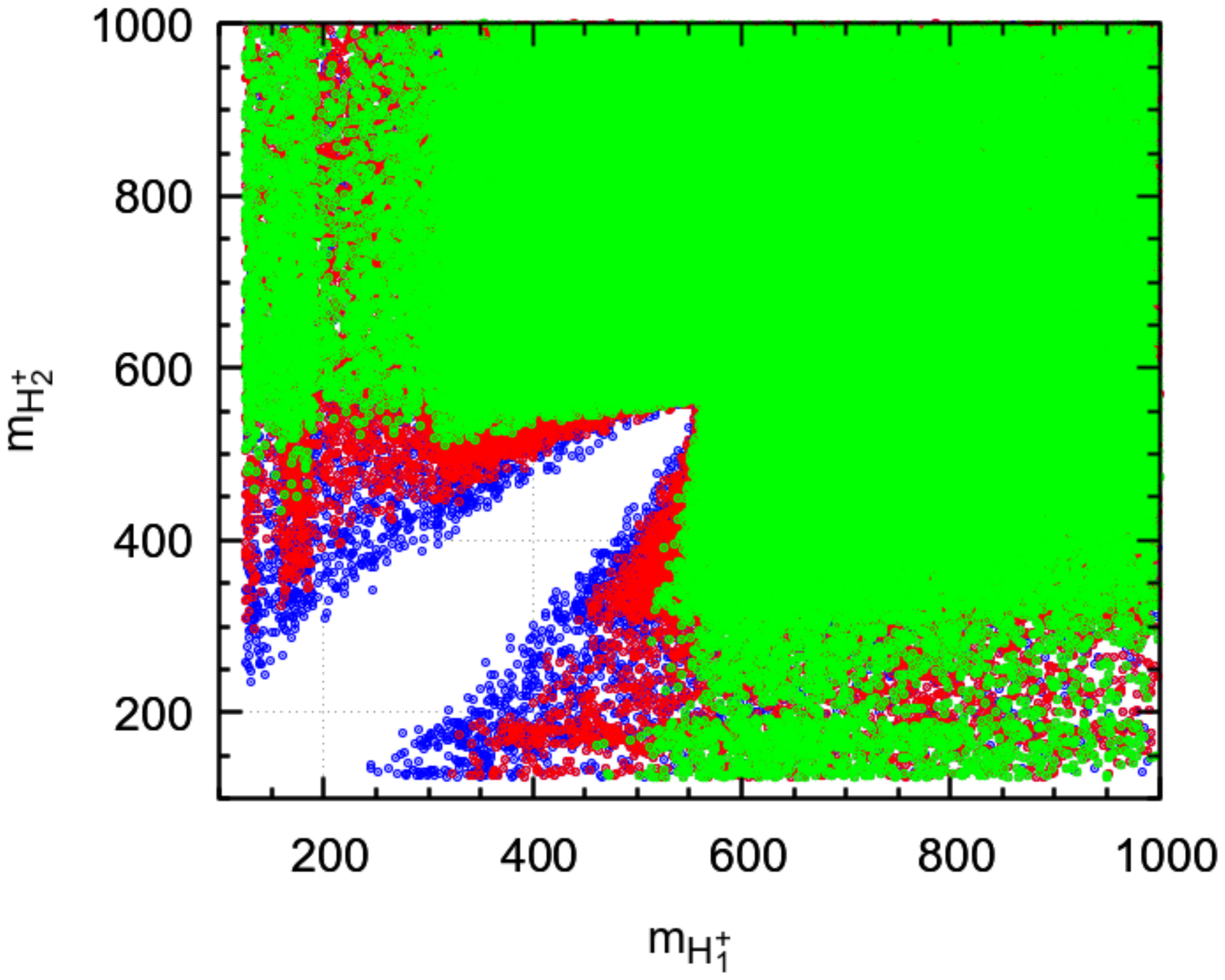}
   \end{tabular}
  \caption{All points are within 1\% of the perfect alignment of
    Eq.~(\ref{eq:4}).
    Left panel: All points passed all constraints except for HB5.
    The blue
    points satisfy Eq.~(\ref{eq:6}). The red points are for
    $\tan\beta_{1,2}>0.5$ and the green points
        are for $\tan\beta_{1,2}>1$. Right panel: same color code as in
    the left panel but only showing points surviving after requiring HB5.}
  \label{fig:6}
\end{figure}

In Fig.~\ref{fig:6} we show the results for the case when we allow the
parameters 
to differ 1\% from the perfect alignment limit.

\begin{figure}[H]
  \centering
  \begin{tabular}{cc}
      \includegraphics[width=0.48\textwidth]{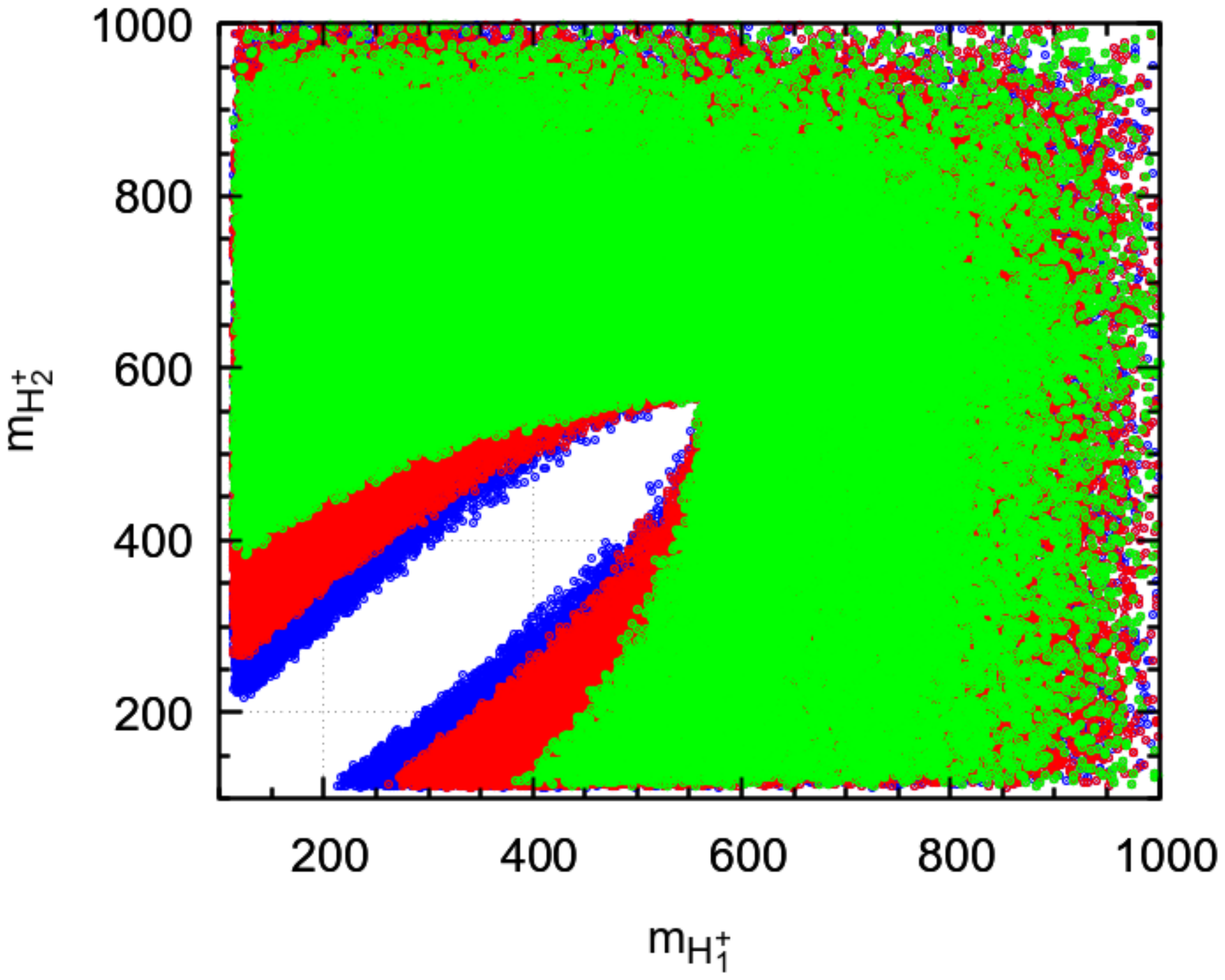}
      &
      \includegraphics[width=0.48\textwidth]{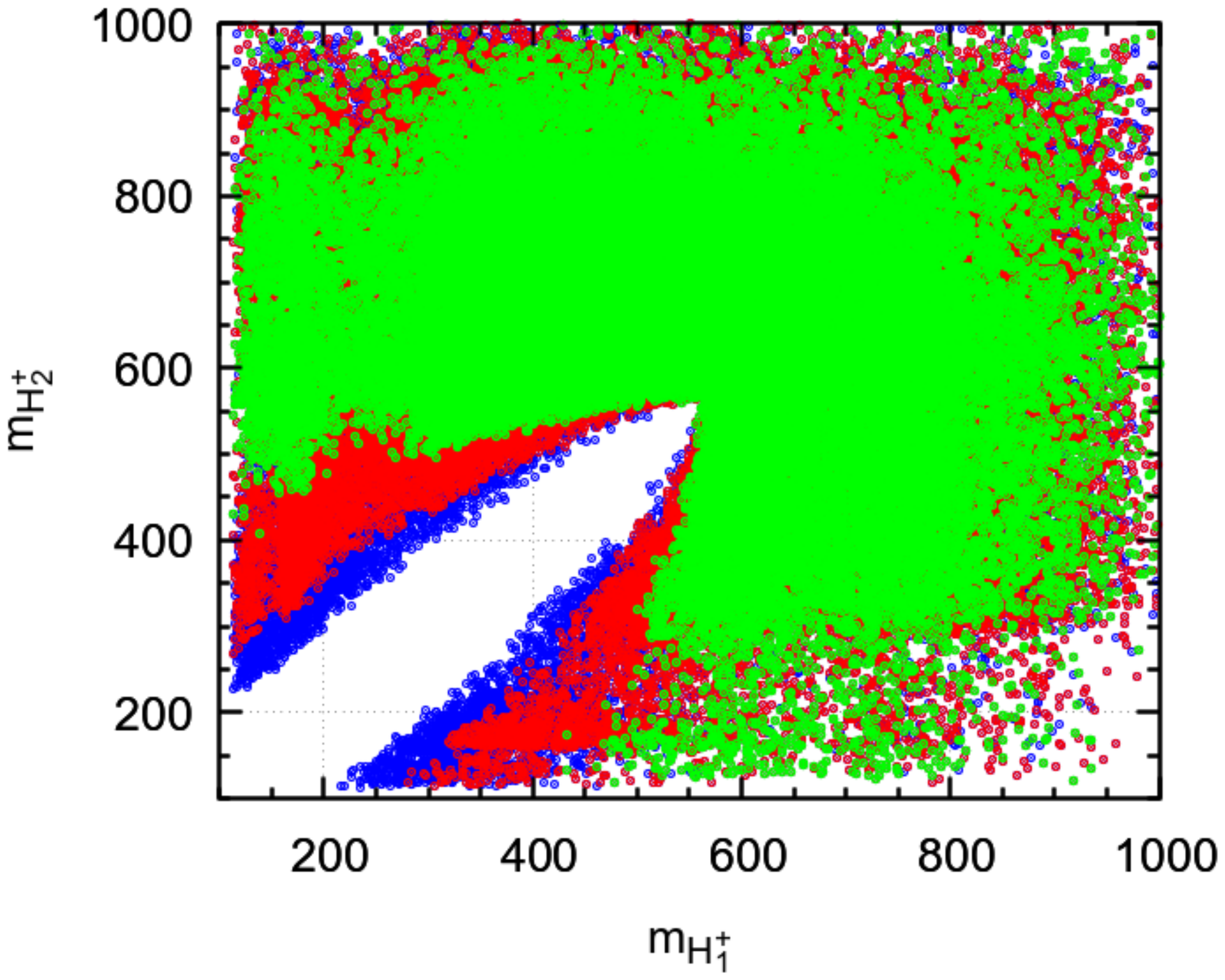}
   \end{tabular}
  \caption{All points are within 10\% of the perfect alignment of
    Eq.~(\ref{eq:4}).
    Left panel: All points passed all constraints except for HB5.
    The blue
    points satisfy Eq.~(\ref{eq:6}). The red points are for
    $\tan\beta_{1,2}>0.5$ and the green points
        are for $\tan\beta_{1,2}>1$. Right panel: same color code as in
    the left panel but only showing points surviving after requiring HB5.}
  \label{fig:7}
\end{figure}

Next we considered the case when the difference for perfect alignment
was 10\%. This is shown in Fig.~\ref{fig:7}.
\noindent
We see that the acceptable points which differ more from perfect alignment
are less frequent, as expected.\footnote{To be more specific,
for the same number of points generated with the constraint of alignment
within 10\% or 1\%, fewer of the former are obtained which pass all
requirements.} 
Nevertheless,
one can still find many points which differ from exact alignment by
as much as 10\%, while satisfying all experimental and theoretical constraints.
And such points do allow for qualitatively different predictions,
as we saw when looking at the charged scalar masses consistent with
$b \rightarrow s \gamma$.
We conclude that imposing perfect alignment is too constraining and
does not cover 
all the interesting features of the $\Z3$ 3HDM.

\subsection{\label{subsec:unusual}Unusual signals of charged scalars}

\begin{figure}[H]
\centering
\includegraphics[width=0.48\textwidth]{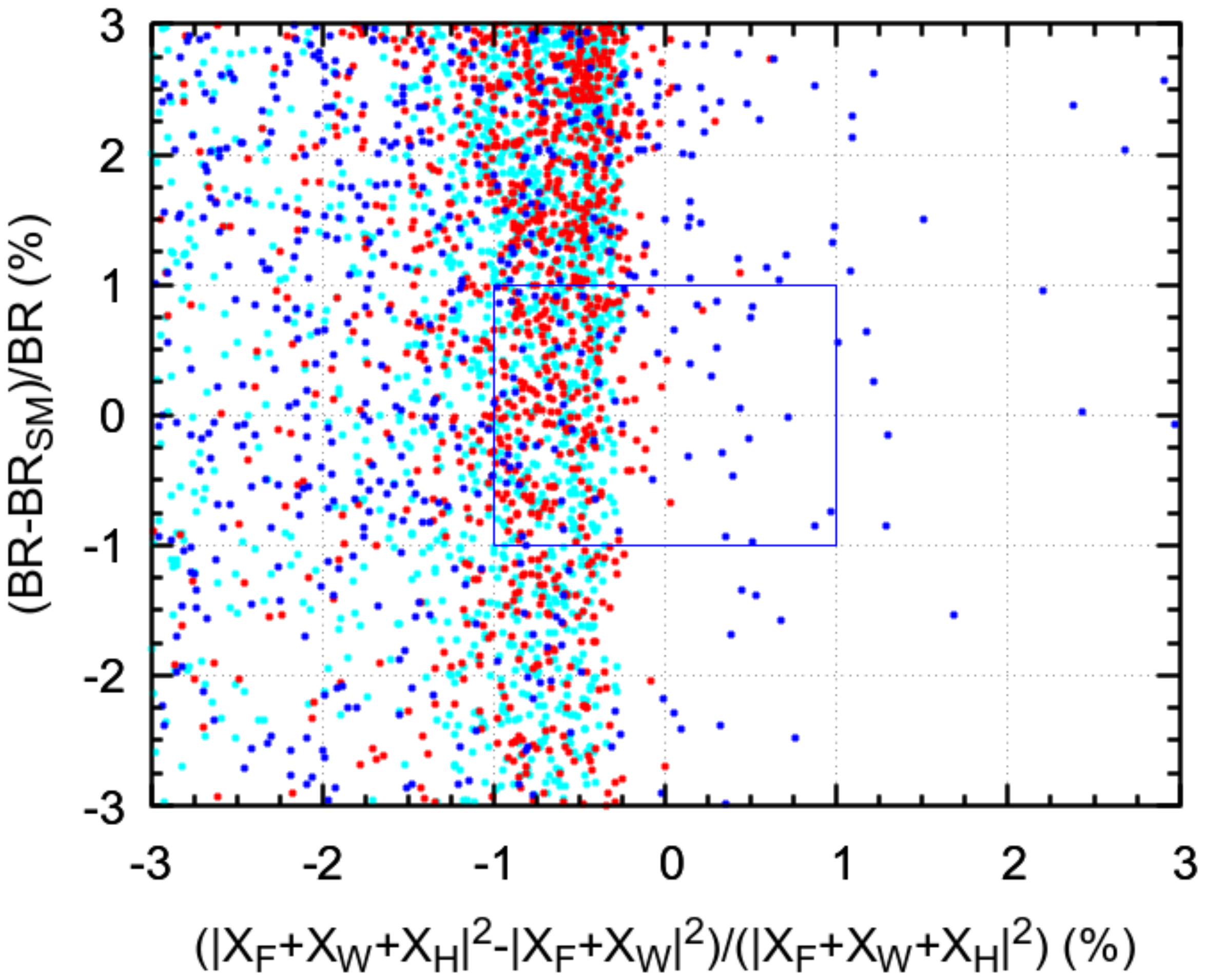}
\caption{\label{fig:cancel} Points with significant approximate
cancellation in both $h \to \gamma \gamma$ (horizontal axis) and
$B \to X_s \gamma$ (vertical),
which pass all theoretical and experimental bounds, including HB5.
Color code: cyan is perfect alignment, red means alignment
within 1\%, and blue means alignment within 10\%.
The blue box guides the eye to those points closest to (0,0).}
\end{figure}

As we have seen,
the contributions of the two charged scalars can exhibit
large cancellations in the decays $h \rightarrow \gamma \gamma$
and $B \rightarrow X_s \gamma$.\footnote{For 3HDMs,
the cancellation
can be exact in $B \rightarrow X_s \gamma$ because there are two
charged components of Higgs doublets feeding the two physical
charged Higgs states. This is no longer the case in the Zee model,
with two Higgs doublets and one charged scalar
singlet \cite{Florentino:2021ybj}.}
For some choices of parameter space, it is even possible that
there are cancellations in both decays simultaneously.
This is illustrated in Fig.~\ref{fig:cancel}.
\noindent
Such charged scalars would, thus, be difficult to probe indirectly.

Notice that points with exact alignment,
in cyan in Fig.~\ref{fig:cancel},
do not allow for cancellation in $h \to \gamma \gamma$;
but alignment with 1\% already does.

Most points within the blue box close to (0,0) have $H_2^+$
decays into quarks or leptons, which  are being sought at LHC.
But there are points which could also be difficult to probe directly with
such common searches, even tough one or both charged scalars
might have relatively small masses.
Indeed,
one can find fine-tuned points in parameter space
where the $H_2^+$ does not decay primordially into quarks or leptons,
but rather as $H_2^+ \rightarrow H_1^+ h_j$ with $h_j=h_1,h_2,A_1$.
We propose that such decays be actively searched for at LHC's next run.
To aid in that experimental endeavour,
we present some benchmark points (BP) in the next section.

\section{\label{sec:BP}Illustrative benchmark points}

This section is devoted to some benchmark points/lines,
with features which may prove useful for the experimental searches.

There has been a recent interest in the literature for unusual decays
of the charged Higgs \cite{Bahl:2021str}, specially those in which the
charged Higgs decays to $W^+ h_i$ where $h_i$ is any of the scalars or
pseudo scalars in the model.

We have performed a search in our large data sets and found many
points where BR($H_1^+\to W^+ + h_{125}$) was larger than 80\%. From
those we selected three benchmark points (BP) that we list in
table~\ref{tab:benchZ3}. 
\begin{table}[htb]
  \centering
  \vspace{-2mm}
  \begin{tabular}{lccc}
    \toprule
    Type-Z & BP1 & BP2 & BP3\\
    \midrule
    $m_{h_2}$ & 419.00 & 494.60  & 486.26 \\
    $m_{h_3}$ & 799.60 & 850.88   & 694.44 \\
    $m_{A_1}$ & 413.80 & 483.96  & 513.46 \\
    $m_{A_2}$ & 763.15 &  806.44 & 647.56 \\
    $m_{H_1^\pm}$ & 396.13 & 477.63  & 506.36 \\
    $m_{H_2^\pm}$ & 752.81 &  843.034  & 654.77 \\
    $(m_{12}^2)$ & -8350 & -31768 & -19562   \\
    $(m_{13}^2)$ & -83278   & -80800 & -63134  \\
    $(m_{23}^2)$ & -231428 & -232361 & -197019  \\
    $\alpha_1$ &   1.289   & 1.343 & 1.328 \\
    $\alpha_2$ & 0.5419 & 0.4406 & 0.7119 \\
    $\alpha_3$ &  0.00543  & -0.00299 &  0.01136 \\
    $\gamma_1$ & -0.00503 & 0.00322  & -0.01078 \\
    $\gamma_2$ & -0.00504 & 0.00301 & -0.01011  \\
    $\beta_1$ &  1.192  & 1.263 & 1.231 \\
    $\beta_2$ & 0.5077   & 0.4311  & 0.7351 \\
    \midrule   
    BR($H_1^+\rightarrow\nu_\tau+\tau^+$) & 0.0688 & 0.0790 &    0.0784 \\
    BR($H_1^+\rightarrow t+\bar{b}$) & 0.0383 & 0.0197 & 0.0358 \\ 
    BR($H_1^+\rightarrow W^+h_1$) & 0.8926 & 0.9011 & 0.8855 \\ 
    \midrule
    BR($H_2^+\rightarrow t+\bar{b}$) & 0.9970 & 0.9995 & 0.9965 \\
    BR($H_2^+\rightarrow W^+h_1$) & 0.0012 & 0.0001 & 0.0009 \\
    BR($H_2^+\rightarrow W^+h_2$) & 0.0007 & 0.0003 & 0.0006 \\
    \bottomrule
  \end{tabular}
  \caption{Benchmark points for the Type Z $Z_3$-3HDM.}
  \label{tab:benchZ3}
\end{table}
For each of these BP we let the mass of the $H_1^+$ vary, leaving
all the other parameters fixed, obtaining benchmark lines. All these
points verify all the constraints, including those from
\texttt{HiggsBounds-5.9.1}. These BP all have the characteristic that
the dominant decay of the charged $H_1^+$ is not in the $tb$ channel,
but in $W^+ h_{125}$, which makes these interesting and deserving to be
searched at the LHC.
\begin{figure}[htb]
\hskip -10mm
  \begin{minipage}{0.7\textwidth}
    \includegraphics[scale=0.4]{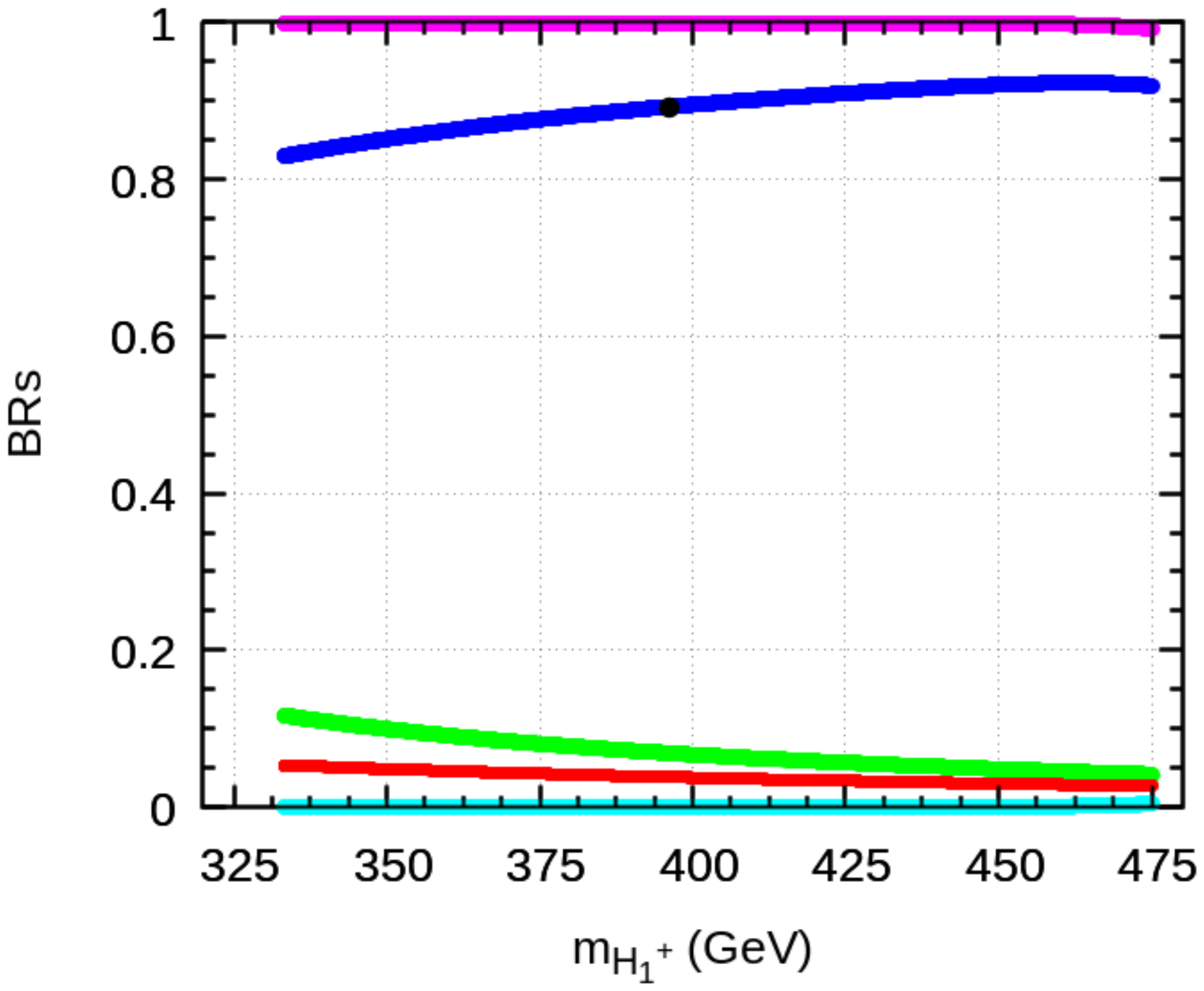}
  \end{minipage}
\hskip -50mm
  \begin{minipage}{0.2\textwidth}
\includegraphics[scale=0.40]{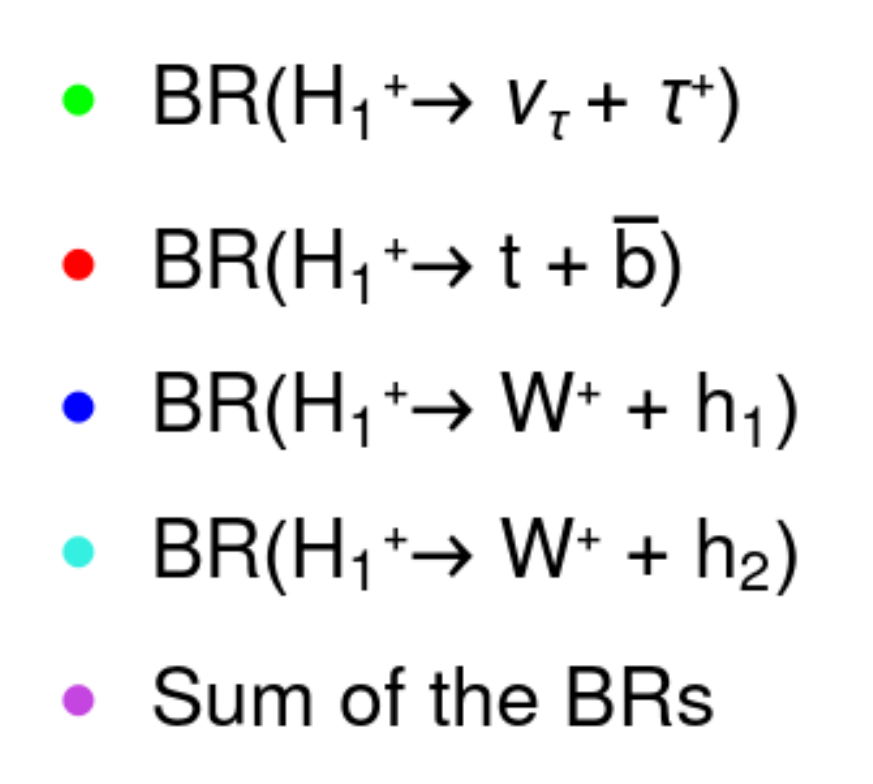}  
\end{minipage}
\caption{\label{fig:P1legend} Most important BR's for BP1. The black
  dot corresponds to the 
original BP in table~\ref{tab:benchZ3}.}
\end{figure}
\begin{figure}[htb]
  \centering
  \hskip -20mm
  \begin{minipage}{0.45\linewidth}
    \includegraphics[scale=0.3]{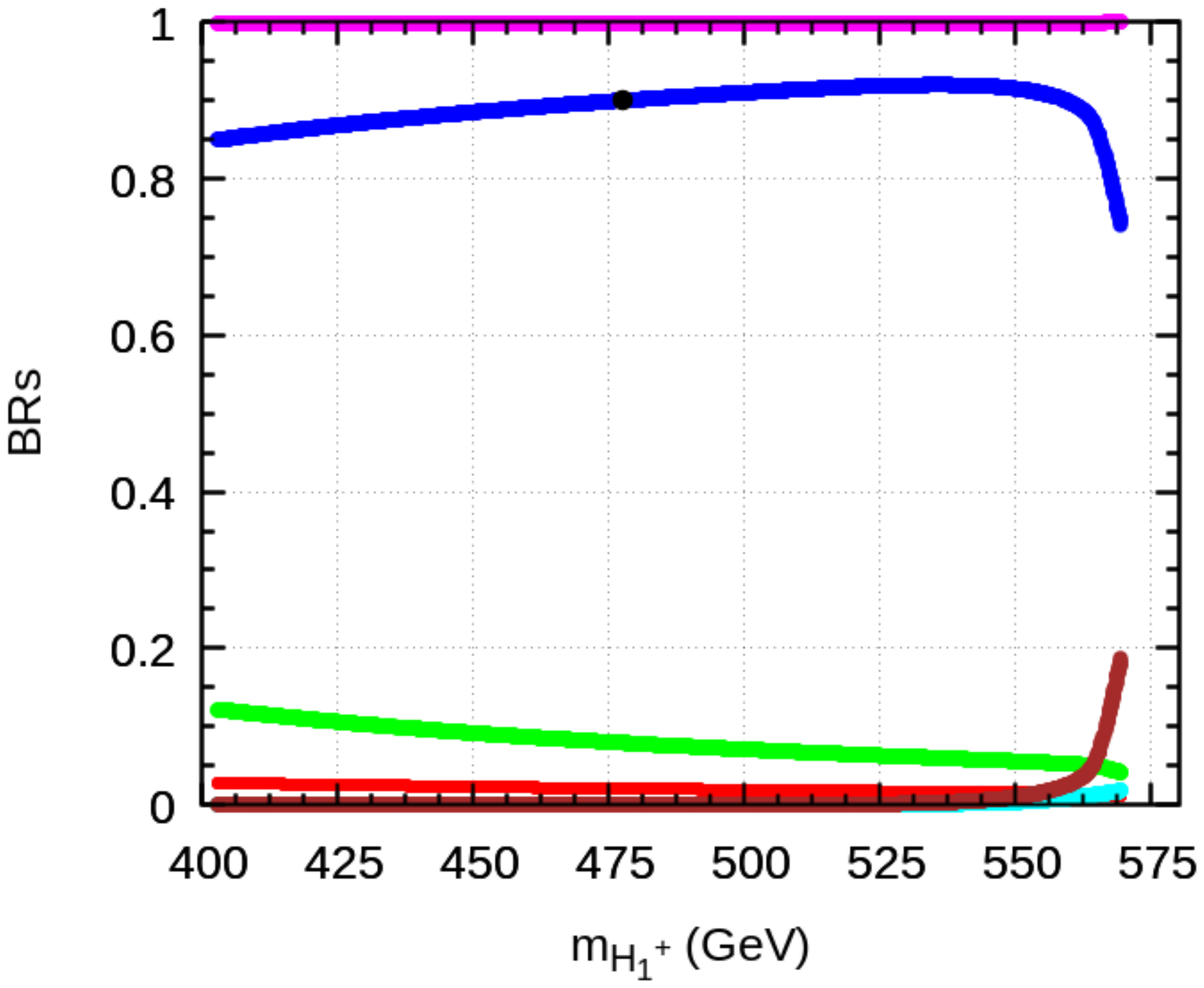}
  \end{minipage}
  \hskip -50mm
  \begin{minipage}{0.2\linewidth}
\includegraphics[scale=0.30]{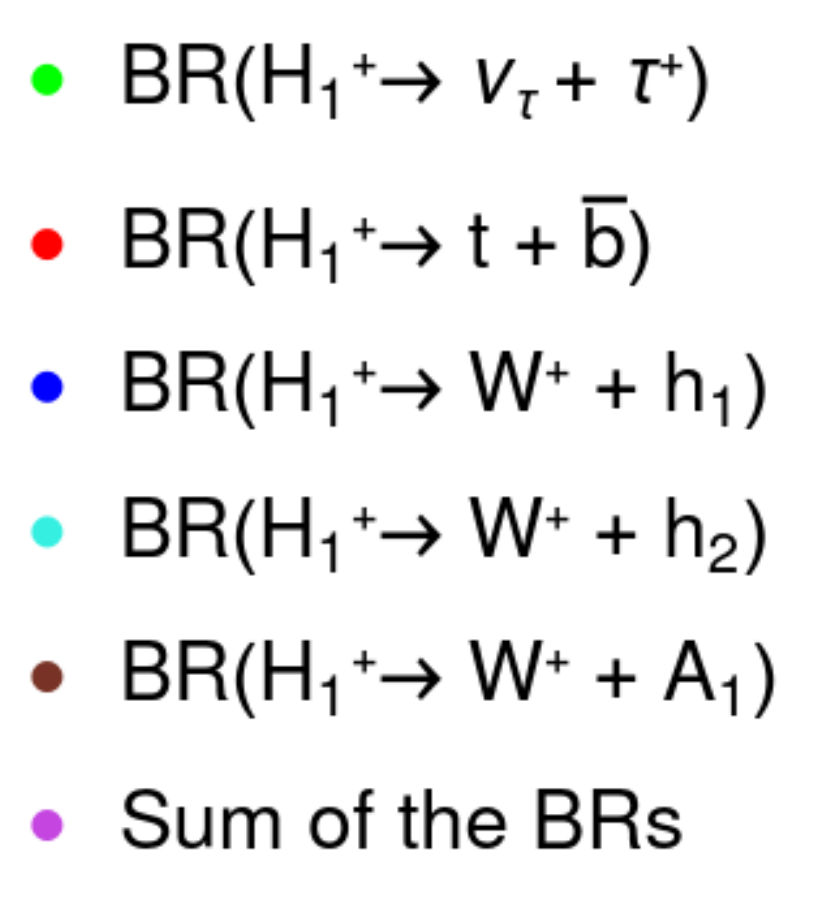}  
\end{minipage}
%
  \hskip 25mm
  \begin{minipage}{0.45\linewidth}
    \includegraphics[scale=0.3]{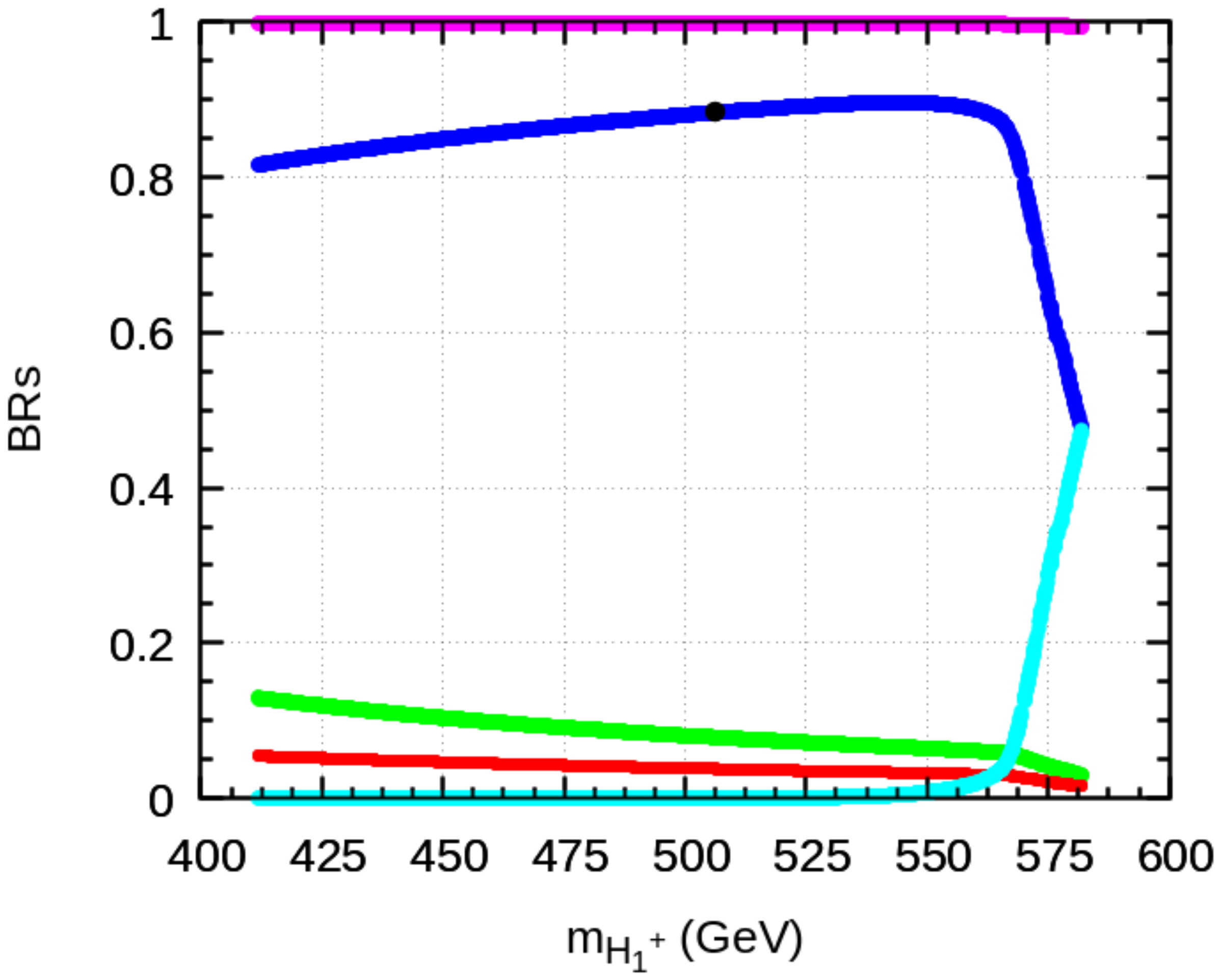}
  \end{minipage}
  \hskip -50mm
  \begin{minipage}{0.2\linewidth}
\includegraphics[scale=0.30]{LegendBP1BP3.pdf}
\end{minipage}
\caption{\label{fig:P3legend} Most important BR's for BP3. The black
  dot corresponds to the 
original BP in table~\ref{tab:benchZ3}.}
\end{figure}
Notice that, for these BP, the other charged Higgs decays 100\% in
$tb$. For BP2 the decay $H_1^+ \to W^+ A_1$ opens up when the mass of
the $H_1^+$ is such that $m_{H_1^+} > M_W + m_{A_1}$ explaining the
decrease in our prefered branching ratio (see
Fig.~\ref{fig:P2legend}). The same happens for the channel  $H_1^+ \to
W^+ h_2$ for BP3 as can be seen in Fig.~\ref{fig:P3legend}.

\section{\label{sec:conclusions}Conclusions}

Multi Higgs models with $N \geq 3$ allow for the possibility that all
fermions of a given charge couple exclusively to one dedicated
scalar. These are known as Type-Z models, and constitute a fifth
alternative beyond the four natural flavour conservation models
allowed in the 2HDM.  We investigate the current bounds on the Type-Z
3HDM imposed by a $\Z3$ symmetry.  We perform an up-to-date analysis
including the latest data for the 125GeV Higgs \cite{Aad:2019mbh},
bounds on new scalars through the \texttt{HiggsBounds-5.9.1} code
\cite{Bechtle:2020pkv}, and the very important theoretical
constraints.

We use the theoretical bounds from unitarity \cite{Bento:2017eti} and
BFB; the latter developed here for the first time.  We stress the
importance of using the most recent LHC bounds, which constrain
severely the allowed parameter space.  In particular, we show that
bounds from $h_2\rightarrow \tau^+ \tau^-$ alter significantly some
results in the literature \cite{Chakraborti:2021bpy}.  This is clearly
visible in our Fig.~\ref{fig:new} and Fig.~\ref{fig:sigBR}.  Moreover,
we also stress the fact that interesting physical observables may
differ significantly when one considers situations close to the
alignment limit, versus adopting the exact alignment limit.  Indeed,
current LHC bounds on the productions and branching ratios of the
125GeV neutral scalar force the measured couplings to lie close to
those obtained for the SM Higgs.  Nevertheless, forcing those
couplings to match \textit{exactly} those in the SM is too
constraining on the parameter space and precludes much of the
interesting new features that the $\Z3$ 3HDM has.

We look at the constraints allowed by current data on the 125GeV Higgs
decays, including a detailed look at $h \to \gamma \gamma$ and its
correlations with the other decays.  We point out the possibility that
the contributions from the two charged scalars might cancel in $h \to
\gamma \gamma$.  This is also possible in $B \to X_s \gamma$, and we
explore explicitly how this allows for lower masses for the charged
scalars.  We provide illustrative benchmark points to aid in
experimental searches.  By comparing the constraints from
\texttt{HiggsBounds-5.7.1} and the newer \texttt{HiggsBounds-5.9.1} we
highlight the importance that the next LHC run will have in further
constraining this model, or perhaps, finally uncovering new physics in
the scalar sector.

\section*{Acknowledgments}
We are very grateful to D. Das for detailed
discussions regarding his Ref.\cite{Chakraborti:2021bpy}.
We are also very grateful to C. Greub for detailed
discussions regarding his Refs.~\cite{Borzumati:1998tg,Borzumati:1998nx}.
JPS is grateful to Z. Ligeti for discussions.
This work is supported in part by the Portuguese
Funda\c{c}\~{a}o para a Ci\^{e}ncia
e Tecnologia\/ (FCT) under Contracts
CERN/FIS-PAR/0008/2019,
PTDC/FIS-PAR/29436/2017,
UIDB/00777/2020,
and UIDP/00777/2020;
these projects are partially funded through POCTI (FEDER),
COMPETE,
QREN,
and the EU.

\appendix

\section{Some important couplings}
\label{appA}

This appendix is devoted to some important couplings
for the $\Z3$ 3HDM used in our calculations.
In our conventions these couplings include the $i$ from the
Feynman rules. These couplings were derived with the help of the
software \texttt{FeynMaster}\cite{Fontes:2019wqh,Fontes:2021iue}.

\subsection{Scalar couplings to $W^\pm$ bosons}

We find for the neutral scalar couplings to $W^+ W^-$,
\begin{align}
&   [h_j,W^+_{\nu},W^-_{\rho}]= i g\, M_W\, g_{\nu\rho}\, 
   \left(\vb{12} {R_{j1}} \, {\hat{v}_1} + {R_{j2}} \, {\hat{v}_2} + {R_{j3}}
       \, {\hat{v}_3} \right) 
\end{align}
Thus,
\be
C_j=R_{j1} \hat{v}_1 + R_{j2} \hat{v}_2 + R_{j3} \hat{v}_3\, ,
\ee
is to be used in Eq.~(\ref{need_Cj}).

\subsection{Scalar couplings to charged Higgs}

The couplings of the scalars $h_j$ with $j=1,2,3$ to the charged Higgs
$H_{k-1}^\mp, H_{l-1}^\pm$ where $k,l=2,3$ (we do not consider here the
charged Goldstone) are,
\begin{align}
     [h_j,H_{k'}^\mp,H_{l'}^\pm]=
     \frac{-i}{2}\, v & 
     \left[\vb{14} 4 \, {\lambda_1} \,
      {Q_{k1}} \, {Q_{l1}} \, {R_{j1}} \, {\hat{v}_1} + 2 \, {\lambda_{5}}
      \, {Q_{k3}} \, {Q_{l3}} \, {R_{j1}} \, {\hat{v}_1} + {\lambda_{7}} \,
      {Q_{k2}} \, {Q_{l1}} \, {R_{j2}} \, {\hat{v}_1} \right.\nonumber\\
    &\left.
      + {\lambda_{10}} \,
      {Q_{k3}} \, {Q_{l1}} \, {R_{j2}} \, {\hat{v}_1} + {\lambda_{7}} \,
      {Q_{k1}} \, {Q_{l2}} \, {R_{j2}} \, {\hat{v}_1} + {\lambda_{11}} \,
      {Q_{k3}} \, {Q_{l2}} \, {R_{j2}} \, {\hat{v}_1}\right.\nonumber\\
  &\left.
      + {\lambda_{10}} \,
      {Q_{k1}} \, {Q_{l3}} \, {R_{j2}} \, {\hat{v}_1} + {\lambda_{11}} \,
      {Q_{k2}} \, {Q_{l3}} \, {R_{j2}} \, {\hat{v}_1} + {\lambda_{10}} \,
      {Q_{k2}} \, {Q_{l1}} \, {R_{j3}} \, {\hat{v}_1}\right.\nonumber\\
  &\left.
      + {\lambda_{8}} \,
      {Q_{k3}} \, {Q_{l1}} \, {R_{j3}} \, {\hat{v}_1} + {\lambda_{10}} \,
      {Q_{k1}} \, {Q_{l2}} \, {R_{j3}} \, {\hat{v}_1} + {\lambda_{12}} \,
      {Q_{k3}} \, {Q_{l2}} \, {R_{j3}} \, {\hat{v}_1}\right.\nonumber\\
  &\left.
      + {\lambda_{8}} \,
      {Q_{k1}} \, {Q_{l3}} \, {R_{j3}} \, {\hat{v}_1} + {\lambda_{12}} \,
      {Q_{k2}} \, {Q_{l3}} \, {R_{j3}} \, {\hat{v}_1} + {\lambda_{7}} \,
      {Q_{k2}} \, {Q_{l1}} \, {R_{j1}} \, {\hat{v}_2}\right.\nonumber\\
  &\left.
      + {\lambda_{10}} \,
      {Q_{k3}} \, {Q_{l1}} \, {R_{j1}} \, {\hat{v}_2} + {\lambda_{7}} \,
      {Q_{k1}} \, {Q_{l2}} \, {R_{j1}} \, {\hat{v}_2} + {\lambda_{11}} \,
      {Q_{k3}} \, {Q_{l2}} \, {R_{j1}} \, {\hat{v}_2} \right.\nonumber\\
  &\left.
      + {\lambda_{10}} \,
      {Q_{k1}} \, {Q_{l3}} \, {R_{j1}} \, {\hat{v}_2} + {\lambda_{11}} \,
      {Q_{k2}} \, {Q_{l3}} \, {R_{j1}} \, {\hat{v}_2} + 4 \, {\lambda_2} \,
      {Q_{k2}} \, {Q_{l2}} \, {R_{j2}} \, {\hat{v}_2} \right.\nonumber\\
  &\left.
      + 2 \, {\lambda_{6}}
      \, {Q_{k3}} \, {Q_{l3}} \, {R_{j2}} \, {\hat{v}_2} + {\lambda_{11}} \,
      {Q_{k2}} \, {Q_{l1}} \, {R_{j3}} \, {\hat{v}_2} + {\lambda_{12}} \,
      {Q_{k3}} \, {Q_{l1}} \, {R_{j3}} \, {\hat{v}_2}\right.\nonumber\\
  &\left.
      + {\lambda_{11}} \,
      {Q_{k1}} \, {Q_{l2}} \, {R_{j3}} \, {\hat{v}_2} + {\lambda_{9}} \,
      {Q_{k3}} \, {Q_{l2}} \, {R_{j3}} \, {\hat{v}_2} + {\lambda_{12}} \,
      {Q_{k1}} \, {Q_{l3}} \, {R_{j3}} \, {\hat{v}_2} \right.\nonumber\\
  &\left.
      + {\lambda_{9}} \,
      {Q_{22}} \, {Q_{l3}} \, {R_{j3}} \, {\hat{v}_2} + 2 \, {\lambda_4} \,
      \left( {Q_{22}} \, {Q_{l2}} \, {R_{j1}} \, {\hat{v}_1} + {Q_{k1}} \,
        {Q_{l1}} \, {R_{j2}} \, {\hat{v}_2} \right) \right.\nonumber\\
  &\left.
      + {\lambda_{10}} \,
      {Q_{22}} \, {Q_{l1}} \, {R_{j1}} \, {\hat{v}_3} 
      + {\lambda_{8}} \,
      {Q_{k3}} \, {Q_{l1}} \, {R_{j1}} \, {\hat{v}_3} + {\lambda_{10}} \,
      {Q_{k1}} \, {Q_{l2}} \, {R_{j1}} \, {\hat{v}_3} \right.\nonumber\\
  &\left.
      + {\lambda_{12}} \,
      {Q_{k3}} \, {Q_{l2}} \, {R_{j1}} \, {\hat{v}_3} + {\lambda_{8}} \,
      {Q_{k1}} \, {Q_{l3}} \, {R_{j1}} \, {\hat{v}_3} + {\lambda_{12}} \,
      {Q_{22}} \, {Q_{l3}} \, {R_{j1}} \, {\hat{v}_3} \right.\nonumber\\
  &\left.
      + {\lambda_{11}} \,
      {Q_{22}} \, {Q_{l1}} \, {R_{j2}} \, {\hat{v}_3} + {\lambda_{12}} \,
      {Q_{k3}} \, {Q_{l1}} \, {R_{j2}} \, {\hat{v}_3} + {\lambda_{11}} \,
      {Q_{k1}} \, {Q_{l2}} \, {R_{j2}} \, {\hat{v}_3}\right.\nonumber\\
  &\left.
      + {\lambda_{9}} \,
      {Q_{k3}} \, {Q_{l2}} \, {R_{j2}} \, {\hat{v}_3} + {\lambda_{12}} \,
      {Q_{k1}} \, {Q_{l3}} \, {R_{j2}} \, {\hat{v}_3} + {\lambda_{9}} \,
      {Q_{22}} \, {Q_{l3}} \, {R_{j2}} \, {\hat{v}_3}\right.\nonumber\\
  &\left.
      + 2 \, {\lambda_{5}}
      \, {Q_{k1}} \, {Q_{l1}} \, {R_{j3}} \, {\hat{v}_3} + 2 \,
      {\lambda_{6}} \, {Q_{22}} \, {Q_{l2}} \, {R_{j3}} \, {\hat{v}_3}
\right.\nonumber\\
  &\left.
      + 4
      \, {\lambda_3} \, {Q_{k3}} \, {Q_{l3}} \, {R_{j3}} \, {\hat{v}_3}
      \vb{14} \right]\nonumber\\
    \hskip -5mm
    \equiv &\, i\, v\, \lambda_{h_j,H_{k'}^+,H_{l'}^-}\, ,
\end{align}
where we have defined $k'\equiv k-1$, $l'\equiv l-1$
with $j=1,2,3$ and $k,l=2,3$. Recall that $\hat{v}_k = v_k/v$. The
coupling $\lambda_{h_j,H_{k'}^+,H_{l'}^-}$ is to be used in Eq.~(\ref{XHformula}).

\subsection{Pseudoscalar couplings to charged Higgs}

The couplings of the pseudoscalars $A_{j'}$ with $j'=j-1$ and $j=2,3$ (we do not
consider the coupling of the neutral Goldstone) are
\begin{align}
  [A_{j'},H_{1}^\mp,H_{2}^\pm]=&\pm
\frac{1}{2} \, v
  \left[ \vb{14} {\lambda_{8}} \, {P_{j3}} \, {Q_{23}}
    \, {Q_{31}} \, {\hat{v}_1} + {\lambda_{11}} \, {P_{j2}} \, {Q_{23}} \,
    {Q_{32}} \, {\hat{v}_1} - {\lambda_{12}} \, {P_{j3}} \, {Q_{23}} \,
    {Q_{32}} \, {\hat{v}_1} \right.\nonumber\\
  &\left.
    - {\lambda_{8}} \, {P_{j3}} \, {Q_{21}} \,
    {Q_{33}} \, {\hat{v}_1} 
  - {\lambda_{11}} \, {P_{j2}} \, {Q_{22}} \,
    {Q_{33}} \, {\hat{v}_1} + {\lambda_{12}} \, {P_{j3}} \, {Q_{22}} \,
    {Q_{33}} \, {\hat{v}_1} \right.\nonumber\\
  &\left.+ {\lambda_{11}} \, {P_{j3}} \, {Q_{22}} \,
    {Q_{31}} \, {\hat{v}_2} - {\lambda_{12}} \, {P_{j3}} \, {Q_{23}} \,
    {Q_{31}} \, {\hat{v}_2}
    - {\lambda_{11}} \, {P_{j3}} \, {Q_{21}} \,
    {Q_{32}} \, {\hat{v}_2} \right.\nonumber\\
  &\left.
    - {\lambda_{11}} \, {P_{j1}} \, {Q_{23}} \,
    {Q_{32}} \, {\hat{v}_2} + {\lambda_{9}} \, {P_{j3}} \, {Q_{23}} \,
    {Q_{32}} \, {\hat{v}_2} + {\lambda_{12}} \, {P_{j3}} \, {Q_{21}} \,
    {Q_{33}} \, {\hat{v}_2}\right.\nonumber\\
  &\left.
    + {\lambda_{11}} \, {P_{j1}} \, {Q_{22}} \,
    {Q_{33}} \, {\hat{v}_2} - {\lambda_{9}} \, {P_{j3}} \, {Q_{22}} \,
    {Q_{33}} \, {\hat{v}_2}\right.\nonumber\\
  &\left.
    + {\lambda_{7}} \,  \left( {Q_{22}} \, {Q_{31}}
      - {Q_{21}} \, {Q_{32}} \right)  \,  \left( {P_{j2}} \, {\hat{v}_1} -
      {P_{j1}} \, {\hat{v}_2} \right)  - {\lambda_{11}} \, {P_{j2}} \,
    {Q_{22}} \, {Q_{31}} \, {\hat{v}_3}\right.\nonumber\\
  &\left.
    - {\lambda_{8}} \, {P_{j1}} \,
    {Q_{23}} \, {Q_{31}} \, {\hat{v}_3} + {\lambda_{12}} \, {P_{j2}} \,
    {Q_{23}} \, {Q_{31}} \, {\hat{v}_3} + {\lambda_{11}} \, {P_{j2}} \,
    {Q_{21}} \, {Q_{32}} \, {\hat{v}_3}\right.\nonumber\\
  &\left.
    + {\lambda_{12}} \, {P_{j1}} \,
    {Q_{23}} \, {Q_{32}} \, {\hat{v}_3} - {\lambda_{9}} \, {P_{j2}} \,
    {Q_{23}} \, {Q_{32}} \, {\hat{v}_3} + {\lambda_{8}} \, {P_{j1}} \,
    {Q_{21}} \, {Q_{33}} \, {\hat{v}_3} \right.\nonumber\\
  &\left.
    - {\lambda_{12}} \, {P_{j2}} \,
    {Q_{21}} \, {Q_{33}} \, {\hat{v}_3} 
    - {\lambda_{12}} \, {P_{j1}} \,
    {Q_{22}} \, {Q_{33}} \, {\hat{v}_3} + {\lambda_{9}} \, {P_{j2}} \,
    {Q_{22}} \, {Q_{33}} \, {\hat{v}_3} \right.\nonumber\\
  &\left.
    + {\lambda_{10}} \,  \left(\vb{11} - 
        {P_{j3}} \, {Q_{22}} \, {Q_{31}} \, {\hat{v}_1}   - {P_{j2}}
      \, {Q_{23}} \, {Q_{31}} \, {\hat{v}_1} 
      + {P_{j3}} \, {Q_{21}} \,
      {Q_{32}} \, {\hat{v}_1} \right.\right.\nonumber\\
  &\left.\left.
      + {P_{j2}} \, {Q_{21}} \, {Q_{33}} \, {\hat{v}_1} +
      {P_{j1}} \, {Q_{23}} \, {Q_{31}} \, {\hat{v}_2}
      - {P_{j1}} \, {Q_{21}}
      \, {Q_{33}} \, {\hat{v}_2} \right.\right.\nonumber\\
  &\left.\left.
      + {P_{j1}} \, {Q_{22}} \, {Q_{31}} \, {\hat{v}_3}
      - {P_{j1}} \, {Q_{21}} \, {Q_{32}} \, {\hat{v}_3}\vb{11} \right)
    \vb{14} \right],   \nonumber\\
    \equiv &\,  v\, \lambda_{A_{j'} H_{1}^\mp H_{2}^\pm}\, ,
\end{align}
for $j'=j-1$ and $j=2,3$. Note that $\lambda_{A_{j'} H_{1}^\mp
  H_{1}^\pm}$ and $\lambda_{A_{j'} H_{2}^\mp H_{2}^\pm}$ vanish.

\providecommand{\href}[2]{#2}\begingroup\raggedright\endgroup


\end{document}